\documentclass[a4paper,pre,twocolumn,showpacs,amsmath,amssymb]{revtex4}

\usepackage{graphicx}
\usepackage{amsmath}

\DeclareMathOperator{\sech}{sech}

\DeclareMathOperator{\Real}{Re}
\DeclareMathOperator{\Imag}{Im}

\newcommand{\conj}[1]{{#1}^{\ast}}

\newcommand{\order}[1]{{\cal O}\left({#1}\right)}

\usepackage[colorlinks,citecolor=blue,linkcolor=blue,filecolor=blue,urlcolor=blue]{hyperref}

\begin{document}

\title{Moving solitons in the discrete nonlinear Schr\"odinger equation}
\author{O. F. Oxtoby}
\email{Oliver.Oxtoby@uct.ac.za}
\author{I. V. Barashenkov}
\email{Igor.Barashenkov@uct.ac.za}
\affiliation{Department of Maths and Applied Maths, University of Cape
Town, Rondebosch 7701, South Africa}
\date{17 July 2007}

\begin{abstract}
Using the method of asymptotics beyond all orders,
we evaluate the amplitude of radiation from a moving small-amplitude
soliton in the discrete nonlinear Schr\"odinger equation.
When the nonlinearity is of the cubic type, this amplitude
is shown to be nonzero for all velocities and therefore small-amplitude
solitons moving without emitting radiation do not exist.
In the case of a \emph{saturable} nonlinearity, on the other hand, 
the radiation is found to be completely suppressed when the soliton moves 
at one of certain
isolated
`sliding velocities'.
We show that a discrete soliton moving at a general speed will
experience radiative deceleration until it either stops and remains
pinned to the lattice, or---in the saturable case---%
locks, metastably, onto one of the sliding velocities.
When the soliton's amplitude is small, however, this deceleration is  extremely slow;  hence,
despite losing energy to radiation, the 
discrete soliton may spend an exponentially long time travelling
with virtually unchanged amplitude and speed.

\end{abstract}


\maketitle

\section{Introduction}

This paper deals with moving solitons of the discrete
nonlinear Schr\"odinger (DNLS) equation.
The earliest applications of the cubic DNLS equation were to the
self-trapping of electrons in lattices (the polaron problem)
and energy transfer in biological chains (Davydov solitons)
-- see the reviews \cite{hennig,braun,scott} for references.
Relatedly, the equation arises in the description of
small-amplitude breathers in Frenkel-Kontorova chains
with weak coupling \cite{braun}.
In optics the equation describes
light-pulse propagation in nonlinear waveguide arrays in the
tight-binding limit \cite{christodoulides,campbell}.
Most recently the DNLS equation has been used to model
Bose-Einstein condensates in optically induced lattices
\cite{trambettoni}.

The question of the existence of moving solitons in the
DNLS equation has been the
subject of debate for some time
\cite{duncan,flach,eilbeck,ablowitz,Kundu,pelinovsky,kevrekidis,
Feddersen,FKZ,Flach_Kladko,
Malomed_collisions_1,Malomed_collisions_2,Malomed_nonlinearity_management}.
Recently, G\'omez-Garde\~nes, Flor\'ia, Falo, Peyrard and Bishop
\cite{gomez1,gomez}
have demonstrated that the stationary motion of pulses in the cubic one-site DNLS
(the `standard' DNLS) is only possible over an oscillatory
 background consisting of a superposition of plane waves.
This result was
obtained by numerical continuation of the moving
Ablowitz-Ladik breather with two commensurate time scales.
In our present paper, we study the travelling discrete solitons analytically
and independently of any reference models.
Consistently with the conclusions of \cite{gomez1,gomez},
we will show that solitons cannot freely move in the cubic DNLS equation;
they emit radiation, decelerate and eventually become
pinned by the lattice.
We shall show, however,
 that this radiation is exponentially small in the soliton's
amplitude, so that broad, small-amplitude pulses are highly mobile and
are for all practical purposes indistinguishable from
freely moving solitons.

In the context of optical waveguide arrays---important
not only in themselves but also as a first step
to understanding more complicated
optical systems such as photonic crystals---%
interest among experimentalists \cite{efremidis,fleischer,chen}
has recently shifted away from media
with pure-Kerr nonlinearity (which gives rise to a cubic term in the DNLS
equation) and towards photorefractive media, which exhibit a
saturable nonlinearity \cite{cuevas,fleischer-nature,
vicencio,hadzievski,stepic}.  In practice such arrays may
be optically induced in a photorefractive material
\cite{efremidis,fleischer} or fabricated permanently -- see
\cite{chen}, for example.
The study of solitons in \emph{continuous} optical
systems with saturable nonlinearitites has a
long history; interesting phenomena here
include
bistability \cite{gatz1,gatz2}, fusion \cite{tikhonenko}
and radiation effects \cite{vidal} which do not arise in
the cubic equation. As for the \emph{discrete} case, the work of
Khare, Rasmussen, Samuelsen and Saxena \cite{khare}
suggests that the saturable one-site
DNLS may be \emph{exceptional} in the sense of ref.\,\cite{BOP};
that is, despite not
being a translation invariant system, it supports translationally
invariant stationary solitons.
This property is usually
seen as a prerequisite for undamped motion in
discrete equations (see e.g.\ \cite{OPB}) and indeed,
the numerical experiments of
Vicencio and Johansson \cite{vicencio} have revealed that soliton
mobility is enhanced in the saturable DNLS equation.

The DNLS equation with a saturable nonlinearity is the second object of
our analysis here; our conclusions will turn out to be in agreement with
the numerical observations of ref.\,\cite{vicencio}.
We will show that for nonlinearities which
saturate at a low enough intensity, solitons can
\emph{slide}---that is, move without
radiative deceleration---at certain isolated velocities.
These `sliding' solitons are examples of embedded solitons.

The usual saturable DNLS equation is the
discrete form of the Vinetskii-Kukhtarev model \cite{vinetskii}:
\begin{equation}
\label{VK}
i\dot{\Phi}_n + \Phi_{n+1} - 2\Phi_n + \Phi_{n-1} -
\gamma\frac{1}{1+|\Phi_n|^2} \Phi_n = 0.
\end{equation}
In order to encompass both cubic and saturable nonlinearities in a single
model, we shall instead consider the equation
\begin{equation}
\label{DNLS}
i\dot{\phi}_n + \phi_{n+1} + \phi_{n-1}
+ \frac{2|\phi_n|^2}{1+\mu|\phi_n|^2}\phi_n = 0,
\end{equation}
obtained from eq.\,\eqref{VK} by making the transformation
$\Phi_n = \sqrt{2/\gamma}e^{-i(2+\gamma)t} \phi_n$ and letting $\mu=2/\gamma$.
In the form above, $1/\mu$ represents the saturation threshold of
the medium \cite{gatz2}, which tends to infinity as one approaches
the pure Kerr (cubic) case of $\mu = 0$.  The higher the value of $\mu$,
the lower is the intensity at which the nonlinearity saturates.

This paper is structured in the following way.
In section \ref{section2},  we construct
a small-amplitude,  broad
travelling pulse  as an asymptotic
series in powers of $\epsilon$, its amplitude. The velocity and frequency of this soliton
are obtained as explicit functions of $\epsilon$  and
its carrier-wave wavenumber.
Then in section \ref{section3}, the main section
of this paper, we derive an expression for
the soliton's radiation tails and measure their amplitude using the method of asymptotics
beyond all orders.  In section \ref{section4}, we investigate the influence of
this exponentially weak radiation on the
soliton's amplitude and speed.  Finally, in  section
\ref{section5}, we summarise our work and make comparisons with some earlier results.

\section{Asymptotic Expansion}
\label{section2}

\subsection{Leading order}
\label{section2_A}

We begin by seeking solutions of the form
\begin{equation}
\label{ansatz}
\phi_n(t) = \psi(X)e^{ikn+i\omega t},
\end{equation}
where
\begin{equation} X = \epsilon(n-vt)
\label{eps_wid}
\end{equation} and $\epsilon$ is a parameter.
By analogy  with the soliton of the continuous NLS,
we expect the discrete
soliton to be uniquely characterised by two parameters---%
e.g.\ $\epsilon$ and $k$---%
while the other two ($\omega$ and $v$)
 are expected to be expressible through
$\epsilon$ and $k$.
Substituting the Ansatz \eqref{ansatz},\,\eqref{eps_wid} into
\eqref{DNLS} gives a differential
advance-delay equation
\begin{multline}
\label{maineqn}
\psi(X+\epsilon)e^{ik}+\psi(X-\epsilon)e^{-ik}-\omega\psi(X) \\
-i\epsilon v\psi'(X) +
\frac{2|\psi(X)|^2} {1+\mu|\psi(X)|^2}  \psi(X) = 0.
\end{multline}
This can be written as an ordinary differential equation of an infinite
order:
\begin{multline}
\label{ode}
 e^{ik}\sum_{n=0}^{\infty}\epsilon^n\frac{1}{n!}\psi^{(n)}
+ e^{-ik}\sum_{n=0}^{\infty}\epsilon^n\frac{(-1)^n}{n!}\psi^{(n)}
-\omega\psi\\-i\epsilon v\psi' +
 \frac{2|\psi|^2} {1+\mu|\psi|^2}  \psi= 0,
\end{multline}
where $\psi^{(n)}= d^n \psi/dX^n$.

From now on we assume that $\epsilon$ is small.
Our aim in this section is to find an approximate solution to
eq.\,\eqref{ode}---and hence eq.\,\eqref{DNLS}---with $\psi={\cal O} (\epsilon)$.
(That is, we are
looking for small, broad
pulses which modulate a periodic carrier wave.)
To this end, we expand
$\psi$, $\omega$ and $v$ as power series in $\epsilon$:
\begin{subequations}
\label{expansions}
\begin{align}
\psi   &= \epsilon(\psi_0 + \epsilon\psi_1 + \cdots),
\label{exp_psi} \\
\omega &= \omega_0 + \epsilon^2 \omega_2 + \cdots,
\label{exp_om}  \\
v      &= v_0 + \epsilon^2 v_2 + \cdots.
\label{exp_v}
\end{align}
\end{subequations}
(We are not expanding $k$ as we consider it, along with $\epsilon$,
as one of the two independent parameters characterising our solution.)
Substituting these expansions into eq.\,\eqref{ode} gives a hierarchy
of equations to be satisfied at each power of $\epsilon$
by choosing  $\omega_n$ and $v_n$ properly.
In nonlinear
oscillations, this perturbation
procedure is known as Lindstedt's method
\cite{Lindstedt}.

At the order $\epsilon^1$ we obtain
\begin{equation}
 \omega_0 = 2\cos k,
 \label{alpha}
 \end{equation}
while  the order $\epsilon^2$ gives
\begin{equation}
v_0=2\sin k.
\label{beta}
\end{equation}
These two relations correspond to the
dispersion of linear waves.
At the power $\epsilon^3$ we obtain the following nonlinear equation
for $\psi_0$:
\begin{equation*}
\cos k \, \psi''_0 - \omega_2\psi_0
+2|\psi_0|^2\psi_0 = 0.
\end{equation*}
This is the stationary form of the NLS equation, which has
the homoclinic solution
\[\psi_0 = a\sqrt{\cos k}\sech(aX)\]
with
\[a^2 = \frac{\omega_2}{\cos k}.\]
Returning to the original variable $\psi$, we note that
 the amplitude $a$ can always be absorbed into
$\epsilon$, the  parameter in eq.\,(\ref{eps_wid}) and
in front of $\psi_0$ in
(\ref{exp_psi}). That is, there is no loss of generality
in setting $a = 1$ and letting $\epsilon$ describe
the amplitude (and inverse width) of the pulse instead.
This allows us to set
\begin{equation}
 \omega_2 = \cos k.
 \label{gamma}
 \end{equation}

 Note that the coefficients in eq.\,\eqref{maineqn} are periodic functions
of the parameter $k$ with period $2 \pi$; therefore it
is sufficient  to consider $k$ in the
interval $(-\pi, \pi)$.
Also, \eqref{maineqn} is invariant with
respect to the transformation $k \rightarrow -k$,
$ \epsilon \rightarrow -\epsilon$, $v \rightarrow -v$;
hence it is sufficient to consider positive $k$
only. Finally, our perturbative solution
does not exist if $\cos k$ is negative.
Thus, from now on we shall assume that $0 \leq k \leq \pi/2$.

\subsection{Higher orders}

At the order $\epsilon^{n+3}$, where $n \ge 1$, we arrive at the following
equations for the real and imaginary parts of $\psi_n$:
\begin{subequations}
\label{genlinear}
\begin{align}
\label{genlinearreal}
{\cal L}_1\Real\psi_n &= \frac{\Real f_{n-1}(X)}{\cos k} \\
\label{genlinearimag}
{\cal L}_0\Imag\psi_n &= \frac{\Imag f_{n-1}(X)}{\cos k},
\end{align}
\end{subequations}
where
\begin{align*}
{\cal L}_0 &= -d^2/dX^2 + 1 - 2\sech^2X, \\
{\cal L}_1 &= -d^2/dX^2 + 1 - 6\sech^2X
\end{align*}
and
\begin{widetext}
\begin{multline}
f_{n-1}(X) = \sum_{j=1}^{[n/2]}\left(\frac{2\cos k}{(2j+2)!}\psi_{n-2j}^{(2j+2)}-\omega_{2j+2}\psi_{n-2j}\right)
+i\sum_{j=1}^{[\frac{n+1}{2}]}\left(\frac{2\sin k}{(2j+1)!}\psi_{n-2j+1}^{(2j+1)}-v_{2j}\psi'_{n-2j+1}\right) \\
+\sum_{m=1}^{n-1}2\psi_0(\psi_m\conj{\psi}_{n-m}-\psi_{n-m}\conj{\psi}_m)
+\sum_{m=1}^{n-1}\sum_{\ell=0}^{n-1-m}2\psi_{n-m-\ell}\psi_m\conj{\psi}_{\ell}\\
+\mu\sum_{m=0}^{n-1}\sum_{\ell=0}^{n-1-m}\Bigg[
\sum_{j=1}^{[\frac{n-m-\ell}{2}]}\left(\frac{2\cos k}{(2j)!}\psi^{(2j)}_{n-m-\ell-2j} -\omega_{2j}\psi_{n-m-\ell-2j}\right)\\
+i\sum_{j=1}^{[\frac{n-m-\ell-1}{2}]}\left(\frac{2\sin k}{(2j+1)!}\psi^{(2j+1)}_{n-m-\ell-2j-1} -v_{2j}\psi'_{n-m-\ell-2j-1}\right)
\Bigg]\psi_m\conj{\psi}_{\ell}.
\label{all_p}
\end{multline}
\end{widetext}
The linear nonhomogeneous ordinary differential equations \eqref{genlinear} must
be solved subject to a boundedness condition.

The bounded homogeneous solutions of eqs \eqref{genlinearreal}   and
\eqref{genlinearimag} ($\sech X\tanh X$ and  $\sech X$, respectively)
correspond to the translation- and U(1)-invariances of eq.\,\eqref{maineqn}.
Including these zero modes in the full solution of eqs \eqref{genlinear}
 would amount just to the translation of $\psi(X)$ by a constant distance in $X$
and its multiplication by a
constant phase factor.  These deformations are trivial,
and hence we can safely discard
the homogeneous solutions at each order of $\epsilon$.

As $\psi_0$'s real part is even and its imaginary  part is odd (zero),
$\psi_1$'s real and imaginary parts are also even and odd, respectively.
The same holds, by induction, to all orders of the perturbation theory.
Indeed, assume
that $\psi_0, \psi_1, \ldots, \psi_{n-1}$ have even real parts and odd
imaginary parts. Then it is not difficult to verify
that the function ${\rm Re} f_{n-1}(X)$ is even and ${\rm Im} f_{n-1}(X)$
is odd.  Since the operators ${\cal L}_0$
and ${\cal L}_1$ are parity-preserving, and since
we have excluded the corresponding homogeneous solutions, this means
that $\psi_n$ has an even real part and an odd imaginary part.
Finally, the homogeneous solution of \eqref{genlinearimag} being even
and that of \eqref{genlinearreal} being odd, the
corresponding solvability conditions are satisfied
at any order.

Note that since the solvability conditions do not impose any constraints
on $v_n$ and $\omega_n$, the coefficients
$v_n$ with $n \geq 2$ and $\omega_n$
with $n \geq 4$ can be chosen completely arbitrarily.

\subsection{Explicit perturbative solution to order $\epsilon^3$}

Solving eqs \eqref{genlinear} successively,
 we can obtain the discrete soliton \eqref{exp_psi} to
any desired accuracy. Here, we restrict ourselves
to corrections up to the cubic power in $\epsilon$.
 The order $\epsilon^3$ is the lowest order
at which the saturation parameter $\mu$ appears in the solution.
On the other hand, it is high enough to exemplify
and motivate
our choice of the coefficients in \eqref{exp_om} and \eqref{exp_v}.

Letting $n=1$ in eq.\,\eqref{all_p}, we have
\begin{equation*}
f_0(X) = \frac{2i}{3!} \sin k \psi_0''' - iv_2\psi'_0.
\end{equation*}
The corresponding solution of
eq.\,\eqref{genlinear} is
\begin{align*}
 \psi_1 = \, &\frac{i}{\sqrt{\cos k}} \Bigg[ \frac{1}{2}\sin k\sech X\tanh X \\
&\;\;\;\;+\frac{1}{2}\left(v_2-\frac{1}{3}\sin k \right)X\sech X \Bigg].
\end{align*}
Here the term proportional to $X\sech X$
decays to zero as $|X| \to \infty$; however, it becomes greater than
$\psi_0(X)$ for sufficiently large $|X|$, leading to
nonuniformity of the expansion \eqref{exp_psi}.
In order to obtain a uniform
expansion,  the term in question
should be eliminated. Being free to choose the coefficients
$v_n$ with $n \geq 2$, we use this freedom to set
\[v_2 = \frac{1}{3}\sin k. \]
 This leaves us with
\[\psi_1 = \frac{i}{2} \sqrt{\cos k}\tan k\sech X\tanh X. \]
After `distilling' in a similar way the correction
 $\psi_2$,
where we fix $$ \omega_4 =  \frac{1}{12} \cos k $$
to eliminate a term proportional to $X \sech X$, we obtain
\begin{align}
\label{regularpert}
\psi &= \epsilon\sqrt{\cos k}\Big\{ \sech X + \frac{i}{2} \epsilon\tan k
\sech X\tanh X \nonumber\\
&\;\;+ \frac{1}{12}\epsilon^2\big[4\sech^3X-3\sech X \nonumber\\
&\;\;\;\;+ \frac{1}{2}\tan^2k(14\sech^3X-13\sech X) \nonumber\\
&\;\;\;\;\;\;+ 4\mu\cos k(2\sech X - \sech^3X)\big]
+ {\cal O}(\epsilon^3)\Big\},
\end{align}
where
\begin{subequations}
\begin{align}
\omega &=
2\left[1+\frac{1}{2!}\epsilon^2+\frac{1}{4!}\epsilon^4+
{\cal O}(\epsilon^6)\right]\cos k,
\label{omegadef} \\
v &=
2\left[1+\frac{1}{3!}\epsilon^2+{\cal O}(\epsilon^4)\right]\sin k.
\label{cdef}
\end{align}
\end{subequations}

\subsection{Velocity and frequency of the discrete soliton}

In the previous subsection we have shown that fixing
suitably the coefficients $\omega_n$ and $v_n$ can lead to a uniform
expansion of $\psi$ to order $\epsilon^3$. Our approach
was based on solving for $\psi_n$ explicitly and then setting
the coefficients in front of the `secular' terms $X \sech X$ to zero.
Here, we show that the secular terms can be eliminated
to all orders -- and
without appealing to explicit solutions.

Assume that the secular terms have been
suppressed in all nonhomogeneous solutions
 $\psi_m$ with $m$ up to $n-1$; that is, let
\begin{align}
\psi_m(X) &\to C_m e^X + o(e^X) \nonumber\\
&\quad\mbox{as} \ X \to -\infty \quad
(m=0,1, \ldots, n-1).
\label{as_m}
\end{align}
(Since we know that the real part of $\psi_m$ is even
and the imaginary part odd, it is sufficient to consider the
asymptotic behaviour at one infinity only.)
The constant $C_m$ may happen to be zero for some $m$
in which case the decay of $\psi_m$ will be faster
than $e^X$.
Our objective is to choose $\omega_n$ and $v_n$ in
such a way that $\psi_n(X)$ will also satisfy \eqref{as_m}.
To this end, we consider the function \eqref{all_p}.
All terms which are trilinear in $\psi_0,\ldots,\psi_{n-1}$
and  derivatives of these functions decay  as $e^{3X}$ or faster; these
terms in $f_{n-1}$
cannot give rise to the secular terms
proportional to $X \sech X$ in
$\psi_n$. On the other hand, the terms making up the first
line in \eqref{all_p} tend to
\begin{align*}
&e^X \sum_{j=1}^{[n/2]} \left[
\frac{2\cos k}{(2j+2)!}
-\omega_{2j+2}\right] C_{n-2j} \\
&\quad +i e^X \sum_{j=1}^{[\frac{n+1}{2}]}\left[
\frac{2\sin k}{(2j+1)!}
-v_{2j}\right] C_{n-2j+1}
\end{align*}
as $X \to -\infty$. These are `asymptotically resonant' terms --
in the sense that
their asymptotics are proportional to the asymptotics  of the
homogeneous solutions of \eqref{genlinearreal} and
\eqref{genlinearimag}.
It is these terms on the right-hand sides
of \eqref{genlinearreal} and
\eqref{genlinearimag} that give rise to the secular terms in the
solution $\psi_n$.
The resonant terms will be suppressed if
we let
\begin{equation}
\omega_{2j+2} = \frac{2 \cos k} { (2 j+2)!}, \quad
v_{2j} =
\frac{2 \sin k} { (2 j+1)!}, \quad
j \ge 1.
\label{om_and_v}
\end{equation}
After the resonant terms have been eliminated, the bounded solution
to \eqref{genlinear} will have  the asymptotic behaviour $\psi_n \to C_n e^X$
as $X \to -\infty$. By induction, this result extends to all $n \ge 0$.

Substituting  \eqref{om_and_v}, together with
\eqref{alpha}--\eqref{gamma}, into
eqs \eqref{exp_om} and \eqref{exp_v}, and summing
up the series, we obtain
\begin{equation}
\omega= 2 \cos k \cosh \epsilon, \quad
v= 2 \sin k \frac{\sinh \epsilon}{\epsilon},
\label{om_and_v_via_eps}
\end{equation}
the frequency and velocity of the discrete soliton
parameterised
in terms of $k$ and $\epsilon$.

Equations \eqref{om_and_v_via_eps}
 coincide with the expressions \cite{AL} of the Ablowitz-Ladik
soliton's velocity and frequency in terms of its amplitude and wavenumber.
The difference between the two sets of answers is in that
eqs \eqref{om_and_v_via_eps} pertain to small-amplitude
solitons only, whereas Ablowitz and Ladik's formulas
are valid for arbitrarily large amplitudes.

Note, also, that the velocity and frequency
\eqref{om_and_v_via_eps}
do not depend on the saturation parameter $\mu$.
This is in contrast to the stationary ($v=0$) soliton of
eqs \eqref{maineqn} and \eqref{ode}
obtained by Khare, Rasmussen, Samuelsen and Saxena \cite{khare}.
The soliton of ref.\,\cite{khare} has its amplitude and
frequency uniquely determined by $\mu$.


\section{Terms beyond all orders of the perturbation theory}
\label{section3}

\subsection{Dispersion relation for linear waves}

As $X \to -\infty$, the series \eqref{exp_psi} reduces to
$\psi(X)= \sum_{n=0}^\infty  \epsilon^{n+1}C_n e^X$.
The convergence of the series $\sum \epsilon^{n+1} \psi_n(X)$
for all $X$ would imply, in particular, the convergence of the
series $\sum \epsilon^{n+1}C_n $. Therefore, if the series
\eqref{exp_psi} converged, the solution $\psi(X)$ would be
decaying to zero as $X \to -\infty$. However, although
we have shown that the series $\sum \epsilon^{n+1} \psi_n(X)$
is asymptotic
to all orders, it does not have to be convergent.
For instance, it is easy to see that the series
cannot converge for $v=\mu=0$ and any $\epsilon$; since the
advance-delay equation \eqref{maineqn} is
translation invariant, this would imply that we have constructed
a family of stationary solitons
with an arbitrary position relative to the lattice.
This, in turn, would contradict the well established fact that the
`standard' cubic DNLS solitons can only be centered on a
site or midway between two adjacent sites \cite{Laedke_Kluth_Spatschek}
(that is, that the `standard' discretisation
of the cubic NLS is not exceptional \cite{pelinovsky-exceptional}).
Thus, we expect that
  the perturbative solution \eqref{exp_psi} satisfies
 $\psi(X) \to 0$ as $|X| \to \infty$ only for some special
choices of $v$, $\epsilon$ and $\mu$.

Can eqs \eqref{maineqn} and \eqref{ode} have a bounded solution
despite the divergence of the corresponding series $\sum
\epsilon^{n+1}C_n $? To gain some insight into this matter, we linearise
eq.\,\eqref{maineqn} about $\psi=0$ and
find nondecaying solutions of the form $\psi = e^{i{\cal Q} X/\epsilon}$
where ${\cal Q}$ is a root of the dispersion relation
\begin{equation}
\label{disprel}
\omega = 2\cos(k+{\cal Q})+ {\cal Q} v.
\end{equation}
It is easy to check that there is at least one such harmonic solution
[i.e.\ eq.\,\eqref{disprel} has at least one root] if $v \neq 0$.
These harmonic waves can form a radiation background over
which the soliton propagates (as suggested by the numerics of
\cite{gomez1,gomez} and the analysis
of a similar problem for the $\phi^4$ kinks in \cite{OPB}). Being
nonanalytic in $\epsilon$, such backgrounds
cannot be captured by any order of the
perturbation expansion.

As we will show below, not only the wavenumbers but also the amplitudes
of the harmonic waves are nonanalytic in $\epsilon$.
This phenomenon was first encountered
in the context of the breather of the continuous
$\phi^4$ model, where Eleonskii, Kulagin, Novozhilova,
and Silin \cite{eleonskii}
suggested that the radiation from
the breather could be
exponentially weak.  Segur and Kruskal \cite{sk,sk2}
then developed the method of `asymptotics beyond all orders' to
demonstrate that, in the limit
$\epsilon \rightarrow 0$, such radiation does exist.
We will use Segur and Kruskal's method to measure
the magnitude of the radiation background of the travelling discrete soliton.

Qualitatively, the fact that the radiation
is not excited at any order of the perturbation expansion
is explained by the fact that the soliton exists on the
long length scale $X$, whereas the radiation has the shorter
scale $X/\epsilon$.  To all orders, the two are uncoupled.

\begin{figure}[btp]
\includegraphics{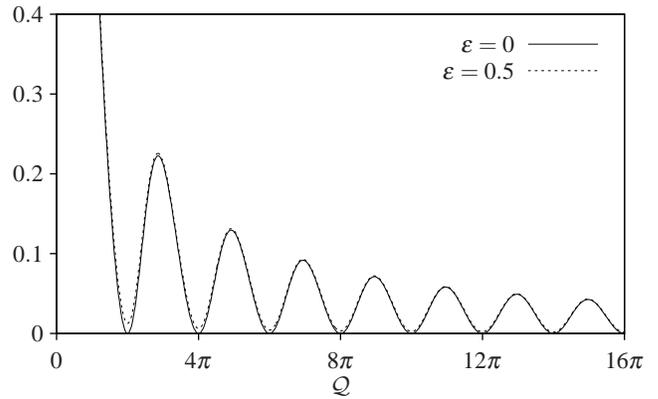}
\caption{\label{nicedisprelfig}The left hand side of eq.\,\eqref{nicedisprel} for
two values of $\epsilon$.  The root(s) ${\cal Q}_n$ of the
dispersion relation \eqref{disprel} are located where this graph
is intersected by a horizontal straight line with ordinate
equal to $\tan k$.}
\end{figure}

Using the relations \eqref{om_and_v_via_eps},
we can rewrite \eqref{disprel} as
\begin{equation}
\label{nicedisprel}
\frac{\cosh\epsilon-\cos {\cal Q}}
{(\sinh\epsilon/ \epsilon){\cal Q} -\sin {\cal Q}} = \tan
k.
\end{equation}
The left hand side is plotted in fig.\,\ref{nicedisprelfig}.  For $\epsilon =
0$ it has minima at multiples of $2\pi$ where the curve is tangent to the horizontal axis.
  For nonzero $\epsilon$, the minima of the curve are lifted off the
${\cal Q}$ axis
slightly.
The minima with larger values of ${\cal Q}$ have smaller elevations above the
horizontal axis; i.e.,
the minima come closer and closer to the ${\cal Q}$
axis as ${\cal Q}$ grows.
We see from the figure that for $k >k^{(1)}_{\rm max}$,
where $k^{(1)}_{\rm max} \approx 0.22$, there is only one
radiation mode. Note also that the left-hand side of
eq.\,\eqref{nicedisprel} is negative for negative ${\cal Q}$; since
we have assumed that $0 \le k \le \pi/2$, this implies
that eq.\,\eqref{nicedisprel} cannot have negative roots.

\subsection{Radiating solitons}
\label{Radi_Soli}

Although we originally constructed the
expansion \eqref{exp_psi} as an asymptotic
approximation to a  solution which is stationary
in the frame of reference moving with the velocity $v$, it can
also represent an approximation to a time-dependent solution
$\psi(X,t)$. Here $\psi(X,t)$ is related to $\phi_n(t)$,
the discrete variable in eq.\,\eqref{DNLS},
by the substitution \eqref{ansatz}:
\begin{equation}
\label{transformation}
\phi_n(t) = \psi(X,t)e^{ikn+i\omega t}.
\end{equation}
  The coefficients $\psi_n$ in the asymptotic expansion of $\psi(X,t)$ will
coincide with the coefficients in the expansion of the stationary
solution $\psi(X)= \sum
\epsilon^{n+1} \psi_n $ if
the time derivatives $\partial_t\psi_n$
lie beyond all orders of $\epsilon$ and hence the time evolution
of the free parameters $k$ and $\epsilon$ occurs on a time scale
longer than
any power of $\epsilon^{-1}$.
Physically, one such  solution represents
a travelling soliton slowing down and attenuating as the Cherenkov
radiation left in its wake carries momentum and energy away from its
core.

Substituting \eqref{transformation} into
eq.\,\eqref{DNLS}, gives
\begin{equation}
\label{timedepeqn}
i\psi_t+\psi^+e^{ik}+\psi^-e^{-ik}-\omega\psi-i\epsilon v\psi_X
+\frac{2|\psi|^2}{1+\mu|\psi|^2} \psi = 0,
\end{equation}
where $\psi^{\pm} = \psi(X\pm\epsilon,t)$.

We consider two solutions of this equation which both have the same
asymptotic expansion \eqref{exp_psi}, denoted $\psi_s(X,t)$
and $\psi_u(X,t)$, such that $\psi_s(X,t)\to 0$ as $X \to +\infty$ and
$\psi_u(X,t) \to 0$ as $X \to -\infty$.
Since the difference $\Psi \equiv \psi_s-\psi_u$ is small
(lies beyond all orders of $\epsilon$),  and since the solution
$\psi_s$ can be regarded as a perturbation of $\psi_u$,
$\Psi$ obeys the linearisation of eq.\,\eqref{timedepeqn}
about $\psi_u$ to a good approximation.  That is,
\begin{multline}
\label{linearised}
i\Psi_t + \Psi^+e^{ik} + \Psi^-e^{-ik} - \omega\Psi
-iv\epsilon\Psi_X \\
+ \frac{4|\psi_u|^2\Psi +2\psi_u^2\conj{\Psi}}{1+\mu|\psi_u|^2}
+ \frac{2\mu|\psi_u|^2(|\psi_u|^2\Psi +\psi_u^2\conj{\Psi})}{(1+\mu|\psi_u|^2)^2} = 0.
\end{multline}
Since $\psi_u = \order{\epsilon}$, we can solve  eq.\,\eqref{linearised}
to leading order in $\epsilon$ by ignoring the last two terms in it; the
resulting solutions are
exponentials of the form $e^{i\mathcal{Q}X/\epsilon-i\Omega t}$. We  make a
preemptive simplification by setting $\Omega = 0$.
[That $\Omega$ has to be set equal to zero follows from
matching these exponentials to the far-field 
asymptotes of the stationary `inner' solution;
 see section \ref{IIID} below.
Physically,  $\Omega = 0$ implies that the travelling pulse will
 only excite the radiation
with its own (zero) frequency in the comoving frame.]
The leading-order solution of eq.\,\eqref{linearised} is therefore
\begin{equation}
\label{stable-minus-unstable-outer}
\Psi = \sum_n A_n e^{i {\cal Q}_nX/\epsilon} + \order{\epsilon^1},
\end{equation}
where ${\cal Q}_n$ ($n=1,2,\ldots$) are the roots, numbered in order from smallest to
largest, of the dispersion relation \eqref{disprel}.
(Recall that since we have taken $k$ in the interval
 $[0, \pi/2]$, all the roots ${\cal Q}_n$ are positive.)
For $k >k_{\rm max}^{(1)} \approx 0.22$, there is only one root, ${\cal Q}_1$.

Higher-order corrections to the solution \eqref{stable-minus-unstable-outer}
can be found
by substituting the ansatz
\begin{multline}
\label{radansatz}
\Psi = \sum_n A_n[1+\epsilon f_1^{(n)}(X)
+\epsilon^2f_2^{(n)}(X)+\cdots]e^{i {\cal Q}_n X/\epsilon} \\
+  \sum_n A_n^* [\epsilon^2 g_2^{(n)}(X)
+\epsilon^3 g_3^{(n)}(X)+\cdots]e^{-i{\cal Q}_n X/\epsilon}
\end{multline}
into eq.\,\eqref{linearised}, expanding the advance/delay terms
$f^{(n)}_{1,2,\ldots}(X\pm\epsilon)$ and
$g^{(n)}_{2,3,\ldots}(X\pm\epsilon)$ in Taylor series in $\epsilon$,
 and making use of
the asymptotic expansion \eqref{exp_psi},\,\eqref{regularpert} for $\psi_u$.
For instance, the first few corrections are found to be
\begin{equation}
\begin{split}
f_1^{(n)}(X) &= \frac{4i\cos k}{2\sin(k+{\cal Q}_n)-v}\tanh X,
\\
g_2^{(n)}(X) &= \frac{2 \cos k}{\omega+ vq_n-2\cos(k-{\cal Q}_n)}\sech^2 X.
\label{f1g2}
\end{split}
\end{equation}

Since $\psi_u \to 0$ as $X \to -\infty$, it follows that
$\psi_s \to \Psi$ as $X \to -\infty$, and hence, once we know
the amplitudes $A_n$, we know the asymptotic behaviour of $\psi_s$.
We shall now employ the method of asymptotics beyond all orders to
evaluate these amplitudes.

\medskip

\subsection{`Inner' equations}

Segur and Kruskal's method allows one to measure the amplitude of
the exponentially small radiation
 by continuing the solution analytically into the complex plane.
 The leading-order term
of $\psi$, $\epsilon\sqrt{\cos k}\sech X$, has singularities at
$X = \frac{i\pi}{2} + i\pi n$, $n=0, \pm 1, \pm 2, \ldots$.
In the vicinity of these points,
the radiation
becomes significant; the qualitative explanation for this is that the
  $\sech$ function forms a sharp spike with a
short length scale near the singularity point,
and hence there is a strong coupling to
the radiation modes, unlike on
the real axis.  The radiation, which is exponentially small
on the real axis, becomes large enough to be measured near the
singularities.

We define a new complex variable $y$, such that $\epsilon y$ is
small in absolute value
when $X$ is near the lowest singularity
in the upper half-plane:
\begin{equation}
\label{inner-variable}
\epsilon y = X-\frac{i\pi}{2}.
\end{equation}
The variables $y$ and $X$ are usually
referred to as the
`inner' and `outer' variables, respectively -- the transformation
to $y$ effectively `zooms in' on the singularity at
$X = \frac{i\pi}{2}$.
We also define $u(y) \equiv \psi(X)$ and $w(y) \equiv \psi^*(X)$.
Continuing eq.\,\eqref{maineqn}
to the complex plane---%
i.e.\ substituting $u(y)$ for $\psi(X)$ and $w(y)$
for $\psi^*(X)$---we obtain, in the limit $\epsilon\to 0$,
\begin{widetext}
\begin{subequations}
\label{inner-equations}
\begin{align}
& e^{ik}u(y+1)+e^{-ik}u(y-1)-2\cos k \, u(y) - 2i\sin k \, u'(y) +
\frac{2u^2w}{1+\mu uw} = 0, \\
& e^{-ik}w(y+1)+e^{ik}w(y-1)-2\cos k \, w(y) + 2i\sin k \, w'(y) +
\frac{2w^2u}{1+\mu wu} = 0.
\end{align}
\end{subequations}
 Here we have used the fact that
 $\omega \to 2\cos k$
and $v \to 2\sin k$ as $\epsilon \rightarrow 0$.  Equations \eqref{inner-equations}
are our `inner
equations'; they are valid in the `inner region' $-\infty< \text{Re}
\, y
< \infty$,
$\text{Im} \, y<0$. (The solution cannot be continued up from the real $X$ axis 
past the singularity at $y = 0$.)

Solving the system
\eqref{inner-equations} order by order,
we can find solutions in the form of
a series in powers of $y^{-1}$.
Alternatively, we can
 make the change of variables \eqref{inner-variable}
in the asymptotic expansion 
\eqref{regularpert} and send
$\epsilon \rightarrow 0$. This gives, for the first few terms,
\begin{subequations}
\label{pertexpy}
\begin{align}
\hat{u} &= \sqrt{\cos k}\left\{ -\frac{i}{y} + \tan k\frac{1}{2y^2}
+ \left[\frac{1}{3}(1-\mu\cos k)+\frac{7}{12}\tan^2k\right]\frac{i}{y^3}
+ \mathcal{O}(y^{-4})\right\}, \\
\hat{w} &= \sqrt{\cos k}\left\{ -\frac{i}{y} - \tan k\frac{1}{2y^2}
+ \left[\frac{1}{3}(1-\mu\cos k)+\frac{7}{12}\tan^2k\right]\frac{i}{y^3}
+ \mathcal{O}(y^{-4})\right\}.
\end{align}
\end{subequations}\end{widetext}
We are using hats over $u$ and $w$ to
distinguish the series solution \eqref{pertexpy}
from other solutions of eq.\,\eqref{inner-equations}
that will appear in the next section. The asymptotic series \eqref{pertexpy}
 may or may not converge.
We note a symmetry $\hat{u}(-y)=-\hat{w}(y)$ of the power-series
solution.

\subsection{Exponential expansion}
\label{IIID}

In order to obtain an expression for the terms which lie beyond
all orders of $y^{-1}$, we substitute $(u,w) = (\hat{u}, \hat{w}) +
(\delta u, \delta w)$ into eqs \eqref{inner-equations}.
Since $\hat{u}$ and $\hat{w}$ solve
the equations to all orders in $y^{-1}$, then provided $\delta u$ and
$\delta w$ are small, they will solve the linearisation of eqs
\eqref{inner-equations} about $(\hat{u},\hat{w})$ for large $|y|$.

Formal solutions to the linearised system  can
be constructed as series in powers of $y^{-1}$.
Because $\hat{u}$ and $\hat{w}$ are both $\order{y^{-1}}$, the
 leading-order expressions for $\delta u$ and $\delta w$ as $y \to
\infty$ are obtained by substituting zero for $\hat{u}$ and $\hat{w}$
in the linearised equations. This gives
\begin{equation}
\begin{split}
\label{leading-order-inner}
\delta u &\to \sum_n J_n \exp(iq_ny), \\
\delta w &\to \sum_n K_n \exp(-iq_ny) \quad \mbox{as} \  y \to \infty,
\end{split}
\end{equation}
where $q_n$ ($n=0,1,2,\ldots$)
are the roots of
\begin{equation}
\label{leading-order-disp}
\cos(k+q)-\cos k + q\sin k = 0.
\end{equation}
Note that the roots $q_n$ with $n \geq 1$ are given by the $\epsilon \to 0$
limits of the roots of
the dispersion equation 
\eqref{nicedisprel}:
$q_n=\lim_{\epsilon \to 0} {\cal Q}_n$. 
In addition, there is a root $q_0=0$ which does not have a $\mathcal{Q}_0$ counterpart.

The full solutions (i.e.\ solutions including corrections
to all orders in $y^{-1}$) will result if we use the full inverse-power series
\eqref{pertexpy}
for $\hat u$ and $\hat w$; these solutions should have the form
\begin{subequations}
\label{expexpansion}
\begin{align}
\label{expexpansiona}
\delta u &= \sum_n K_n\sum_{m=1}^{\infty}
\frac{d_m^{(n)}}{y^m} \exp(-iq_ny), \\
\label{expexpansionb}
\delta w &= \sum_n K_n\left[1+\sum_{m=1}^{\infty}
\frac{c_m^{(n)}}{y^m}\right] \exp(-iq_ny).
\end{align}
\end{subequations}
Note that we have  excluded the terms proportional to $e^{iq_ny}$
from this ansatz (i.e.\ set the amplitudes $J_n$ to zero) as they would become
exponentially large
on the real $X$ axis. [One can readily verify this by making the change of
variables \eqref{inner-variable} in eq.\,\eqref{leading-order-inner}.]
The coefficients $c_1^{(n)}$, $c_2^{(n)}$, \ldots and $d_1^{(n)}$,
$d_2^{(n)}$, \ldots
are found recursively when the ansatz \eqref{expexpansion}
is substituted into the linearised
equations and like powers of $y^{-1}$ collected.
In particular, the first few coefficients are
\begin{subequations}
\begin{align}
c_1^{(n)} &= -\frac{2i\cos k}{\sin(k+q_n)-\sin k}, \nonumber\\
\label{c1c2}
c_2^{(n)} &= \frac{[\cos(k+q_n)-2\cos k]\cos k}{[\sin(k+q_n)
-\sin k]^2}
\end{align}
and
\begin{equation}
d_1^{(n)} = 0, \quad d_2^{(n)} = \frac{\cos k}{\cos(k-q_n)
-\cos k-q_n\sin k}, 
\label{d1n}
\end{equation}
\end{subequations}
where $n=1,2, \ldots$.

Having restricted ourselves to considering the linearised
equations for $\delta u$ and $\delta w$, we have
only taken into account the simple harmonics in
\eqref{leading-order-inner} and \eqref{expexpansion}.
Writing $\delta u= \varepsilon^1 U_1+ \varepsilon^2 U_2+\cdots$ and
 $\delta w=\varepsilon^1 W_1+ \varepsilon^2 W_2+\cdots$, where
 $\varepsilon$ is an auxiliary small parameter
 (not to be confused with our `principal' small parameter $\epsilon$);
 substituting $u=\hat{u}+\delta u$ and
 $w=\hat{w}+\delta w$  in eqs \eqref{inner-equations},
 and solving order-by-order the resulting hierarchy
 of nonhomogeneous linear equations, we can recover all nonlinear
 corrections to $\delta u$ and $\delta w$.
 The $\varepsilon^2$-corrections
will be proportional to
$e^{-i(q_n+q_m)y}$;
higher-order corrections will introduce harmonics
 with higher combination wavenumbers. Later in this section
 it will become clear that $\varepsilon$ is actually of the order $\exp(-\pi
 \mathcal{Q}_1 / 2 \epsilon)$ [see eq.\,\eqref{symb} below]; hence the
 amplitudes of the combination harmonics  will
 be exponentially smaller than that of $\exp(-i q_1 y)$.

Now we return to the object that is of ultimate interest to us in
this work -- that is, to the function $\Psi$ of section \ref{Radi_Soli}
representing the radiation of the moving soliton. We
need to match $\Psi$ to the corresponding object in the
inner region. To this end, we recall that $\Psi=\psi_s-\psi_u$, where
$\psi_s$ and $\psi_u$ are two solutions of the outer equation
\eqref{maineqn} which share the same asymptotic expansion
to all orders. In the limit $\epsilon \to 0$, the
corresponding functions
\begin{equation}
\begin{aligned}
 u_s(y,t) &\equiv \psi_s(X,t),
&w_s(y,t) &\equiv \psi_s^*(X,t), \\
 u_u(y,t) &\equiv \psi_u(X,t),
&w_u(y,t) &\equiv \psi_u^*(X,t) \label{continua}
\end{aligned}
\end{equation}
solve eqs \eqref{inner-equations}
and share the same inverse-power expansions. We express this fact by writing
\begin{align*}
 u_s(y,t) &\sim \hat{u}(y),\quad
&w_s(y,t) &\sim \hat{w}(y),\qquad \\
 u_u(y,t) &\sim \hat{u}(y),\quad
&w_u(y,t) &\sim \hat{w}(y).\qquad
\end{align*}
\begin{widetext}
Therefore, the difference $u_s-u_u$ (which
results from the analytic continuation of the function $\Psi$)
can be identified with $\delta u$ and $w_s-w_u$ with $\delta w$.
Continuing eq.\,\eqref{radansatz} and its
complex conjugate gives,
as $\epsilon \to 0$,
\begin{subequations} \label{cont12}
\begin{align} u_s-u_u = & \sum_n
\lim_{\epsilon \to 0}
A_n(\epsilon) \exp \left(-\frac{\pi \mathcal{Q}_n}{2 \epsilon} \right)
\left[ 1+\order{\frac1y} \right]
\exp(iq_ny) \nonumber\\
&- \sum_n
\lim_{\epsilon \to 0}
A_n^*(\epsilon) \exp \left(\frac{\pi \mathcal{Q}_n}{2 \epsilon} \right)
\left[\frac{2 \cos k}{\omega + vq_n- 2 \cos(k-q_n)}
\frac{1}{y^2}+\order{\frac{1}{y^3}}\right]
\exp(-iq_ny)
\label{cont1}
\end{align}
and
\begin{align}
w_s-w_u = &\sum_n
\lim_{\epsilon \to 0}
A_n^*(\epsilon) \exp \left(\frac{\pi \mathcal{Q}_n}{2 \epsilon} \right)
\left[ 1+\order{\frac1y} \right]
\exp(-iq_ny) \nonumber\\
&- \sum_n
\lim_{\epsilon \to 0}
A_n(\epsilon) \exp \left(-\frac{\pi \mathcal{Q}_n}{2 \epsilon} \right)
\left[\frac{2 \cos k}{\omega + vq_n- 2 \cos(k-q_n)}
\frac{1}{y^2}+\order{\frac{1}{y^3}}\right]
\exp(iq_ny).
\label{cont2}
\end{align}
\end{subequations}
\end{widetext}
[Here we have used eq.\,\eqref{f1g2}.]
Matching  \eqref{cont1} to \eqref{expexpansiona} and \eqref{cont2}
to \eqref{expexpansionb} yields then $K_0=0$ and
\begin{subequations}
\begin{eqnarray}
\lim_{\epsilon \to 0} A_n(\epsilon) \exp\left(-\frac{\pi \mathcal{Q}_n}{2
\epsilon} \right)=0,
\label{Anex1} \\
\lim_{\epsilon \to 0} A_n^*(\epsilon) \exp\left(\frac{\pi \mathcal{Q}_n}{2
\epsilon} \right)=K_n
\label{Anex2}
\end{eqnarray}
\end{subequations}
for $n=1,2, \ldots $. We note that eq.\,\eqref{Anex1} follows from eq.\,\eqref{Anex2},
while the latter equation can be written, symbolically,
as
\begin{equation}
A_n(\epsilon) \longrightarrow
K_n^* \exp\left(-\frac{\pi \mathcal{Q}_n}
{2\epsilon} \right) \quad \text{as} \ \epsilon \to 0.
\label{symb}
\end{equation}
Our subsequent efforts will focus on the evaluation of the
constants $K_n$.

For $k$ greater than $k^{(1)}_{\text{max}}$ (approximately $0.22$),
there is only one radiation
mode and therefore only one pre-exponential factor, $K_1$. For smaller
$k$ we note that the
 amplitude of the $n$th radiation mode, $A_n$,
is a factor of $\exp\left\{
\frac{\pi}{2\epsilon} (\mathcal{Q}_n-\mathcal{Q}_1)\right\}$
smaller than $A_1$, the amplitude of the first mode.
Referring to fig.\,\ref{nicedisprelfig}, it is clear that for $n \geq 3$,
the difference
$\mathcal{Q}_n-\mathcal{Q}_1$ will be no smaller than $\pi$.
As for the second mode, it becomes as significant as the first one
only when $k=\order{\epsilon^2}$ in which case
$(\mathcal{Q}_2-\mathcal{Q}_1)/\epsilon =\order{1}$.
But in our asymptotic expansion of section
\ref{section2}  we assumed, implicitly, that
$k$ is of order $1$ and so the case of $k =\order{\epsilon^2}$
is beyond the scope of
our current analysis. Therefore, for our purposes all
the radiation modes with
$n \geq 2$ (when they exist)
will  have
 negligible amplitudes compared to that of the first mode,
provided $K_1$  is nonzero and $\epsilon$ is small.
For this reason, we shall
only attempt to evaluate $K_1$ in this paper.

\subsection{Borel summability of the asymptotic series}
\label{BLT}

Pomeau,  Ramani, and Grammaticos~\cite{pomeau}
have shown that the radiation
can be measured using the technique of
Borel summation rather than by solving differential equations
numerically,
as in Segur and Kruskal's original approach.  The method has been refined by
(among others)
Grimshaw and Joshi \cite{GrimshawJoshi, GrimshawNLS} and Tovbis,
Tsuchiya,  Jaff\'e
and Pelinovsky \cite{Tovbis1, Tovbis2, Tovbis3, TovPel}, who have applied it to
difference equations.  Most recently it has been applied to
differential-difference equations
in the context of moving kinks in $\phi^4$ models \cite{OPB}.
This is the approach that we will be pursuing here.

Expressing $u(y)$ and $w(y)$ as Laplace transforms
\begin{subequations}
\label{laplace}
\begin{equation}
u(y) = \int_{\gamma}U(p)e^{-py}dp
\end{equation}
and
\begin{equation}
\quad w(y) = \int_{\gamma}W(p)e^{-py}dp,
\end{equation}
\end{subequations}
where $\gamma$ is a contour extending
from the origin to infinity in the complex $p$ plane,
the inner equations \eqref{inner-equations} are cast in the
form of integral equations
\begin{subequations}
\label{integral-equations}
\begin{gather}
f(p)U + \mu [f(p)U]\ast U\ast W + U\ast U\ast W = 0,
\label{fpU} \\
f(-p)W + \mu [f(-p)W]\ast W\ast U + W\ast W\ast U = 0,
\label{gpU}
\end{gather}
\end{subequations}
where
\[
f(p) = (\cosh p - 1)\cos k - i(\sinh p - p)\sin k.
\]
The asterisk $\ast$ denotes the convolution integral,
\[ U(p) \ast W(p) = \int_0^p U(p-p_1) W(p_1) d p_1, \]
where the integration is performed from the origin to the point $p$ on the
complex plane, along the contour $\gamma$.
In deriving eqs \eqref{integral-equations}, we
have made use of the convolution theorem for the Laplace transform
of the form \eqref{laplace}, where the integration is over
a contour in the complex plane rather than a positive
real axis. The theorem states that 
\begin{equation}
u(y) w(y)= \int_\gamma [ U(p) * W(p) ] e^{-py} dp.
\label{conv_thm}
\end{equation}
The proof of this theorem is provided in appendix \ref{A}.

We choose the contour so that $\arg p \to \pi/2$ as $|p| \to
\infty$ along $\gamma$. In this case we have
$e^{-py} \to 0$ as $|p| \to \infty$ for all $y$ along any line
$-\infty< \text{Re} \, y < \infty$ with $\text{Im} \, y<0$.
Therefore, the integrals in \eqref{laplace} converge
for all $y$ along this line
and any bounded $U(p)$ and $W(p)$.

The function $U(p)$ will have singularities
 at the points
where $f(p)$ vanishes  while the
sum of the double-convolution terms in \eqref{fpU} does not. Similarly,
$W(p)$ will have a singularity wherever $f(-p)$ vanishes
 [while the
sum of the double-convolution terms in \eqref{gpU} does not].
Therefore,  $U$ and $W$ may have  singularities at the points where
\begin{equation}
\label{singularities}
\cosh p - 1 = \pm i \tan k \, (\sinh p - p),
\end{equation}
with the top and bottom signs referring to $U(p)$ and $W(p)$,
respectively.
The imaginary roots of eq.\,\eqref{singularities}
with the top sign are at $p = -iq_n$
and those of eq.\,\eqref{singularities}
with the bottom sign at $p = iq_n$,
where $q_n$ are the real roots of
\eqref{leading-order-disp}.
(We remind the reader that all roots $q_n$
are positive.) The point $p = 0$ is not a singularity
as both double-convolution terms in each line of
\eqref{integral-equations} vanish here.
There is always at least one pure imaginary root of eq.\,\eqref{singularities}
(and \emph{only} one if $k > k^{(1)}_{\rm max}$, where $k^{(1)}_{\rm max} \approx 0.22$).

In addition,
there are infinitely many complex roots.  The complex singularities of $U$
[complex roots of the top-sign equation in \eqref{singularities}] are at the
intersections of the curve given by
\begin{subequations}
\begin{equation}
q =  \frac{\cosh \kappa}{\sin k}
\sqrt{1-\sin^2 k \left( \frac{\kappa}{\sinh \kappa} \right)^2 } -\cot k
\label{curve1a}
\end{equation}
with the family of curves described by
\begin{align}
\label{curve2a}
q &= k - {\rm arcsin} \, \left( \frac{\kappa}{\sinh \kappa} \sin k \right)
+ 2 \pi n, \\
&\quad n=1,2, \ldots \, ,\nonumber
\end{align}
\end{subequations}
and at the
intersections of the curve
\begin{subequations}
\begin{equation}
q =  -\frac{\cosh \kappa}{\sin k}
\sqrt{1- \sin^2 k \left(\frac{\kappa}{\sinh \kappa} \right)^2 } -\cot k
\label{curve1b}
\end{equation}
with the family of curves described by
\begin{align}
\label{curve2b}
q &= k + {\rm arcsin} \, \left( \frac{\kappa}{\sinh \kappa} \sin k \right)
+  \pi (2n+1), \\
&\quad n=-2,-3,-4, \ldots \, .\nonumber
\end{align}
\end{subequations}
Here $\kappa$ and $q$ are the real and imaginary part of $p$:
$p=\kappa+iq$.
The curve \eqref{curve1a} looks like a  parabola
opened upwards, with the vertex at $\kappa=q=0$, and the
curve \eqref{curve1b} like a  parabola opened
downwards, with the vertex at $\kappa=0$, $q=-2 \cot k$. The curves
 \eqref{curve2a} and \eqref{curve2b},
on the other hand, look like parabolas for small $\kappa$
but then flatten out
and approach horizontal straight lines as
$\kappa \to \pm \infty$.
[In compiling the list of  these `flat' curves in \eqref{curve2a} and
\eqref{curve2b}, we have taken into account
 that the curve \eqref{curve2a} with $n=0$ does not have
any intersections with the parabola \eqref{curve1a}, and
the curve \eqref{curve2b} with $n=-1$ does not have
any intersections with the parabola \eqref{curve1b}.]
  As $k$ is reduced, the vertex of the
parabola
\eqref{curve1b} moves down along the $q$ axis;  the
intersections of this parabola with the `flat' curves
\eqref{curve2b} approach, pairwise, the $q$ axis. After colliding on
the $q$ axis, pairs of complex roots move away from
each other along it.
[The parabola \eqref{curve1a}
 does not move as $k$ is reduced, but only steepens,
which results in the singularities in the upper half-plane approaching the
imaginary axis but not reaching it until $k = 0$.]
In a similar way, the complex singularities
of $W(p)$ move onto the imaginary axis as $k$ is decreased.
In section \ref{BLT} below we will use the fact that the
distance from any complex
singularity to the origin is larger than $2 \pi$;
this follows from the observation that
the closest points of the curves \eqref{curve2a} and
\eqref{curve2b} to the origin are their intersections with
the $q$ axis. These are further away than $2 \pi$ from the origin.

In addition to singularities at $p= -iq_n$, the function $U(p)$ will
have singularities at points $p=iq_n$, $n=1,2,\ldots$. These are
induced by the cubic terms in eq.\,\eqref{fpU}; for instance,
the singularity at $p=iq_1$ arises from the convolution of
the term proportional to $p$ in $U*U$ with
the function $W$ which has a
singularity at $p=iq_1$. [That $U(p)$ has singularities at
$p=iq_n$ can also be seen directly from eq.\,\eqref{expexpansiona}.]
Similarly, the function $W(p)$ will
have singularities at points $p=-iq_n$, $n=1,2,\ldots$.
By virtue of the nonlinear terms there will also be singularities
at the `combination points' $i(\pm q_n \pm q_m)$, $i(\pm q_n \pm q_m \pm
q_j)$, etc.

 The formal inverse-power series \eqref{pertexpy}, which
we can write as
\begin{equation}
\hat{u}(y)= \sum_{\ell=0}^\infty \rho_\ell \, \frac{\ell !}{y^{\ell+1}},
\quad
\hat{w}(y)= \sum_{\ell=0}^\infty \nu_\ell \,
\frac{\ell !}{y^{\ell+1}},
\label{uywy}
\end{equation}
result from
the Laplace transformation of power series for $U(p)$ and $W(p)$:
\begin{equation}
U(p)= \sum_{\ell=0}^\infty \rho_\ell \, p^\ell,
\quad
W(p)= \sum_{\ell=0}^\infty \nu_\ell \, p^\ell.
\label{UpWp}
\end{equation}
The series \eqref{UpWp} converge in the disk of radius
$q_1$, centred at the origin, and hence can be integrated term by term
only over the portion of the contour $\gamma$ which lies within that disk.
However, by Watson's lemma, the remaining part of the contour
makes an exponentially small contribution to the integral and
the resulting series \eqref{uywy} are asymptotic as $y \to \infty$.
The functions
$u(y)$ and $w(y)$ defined by eqs \eqref{laplace} give the Borel
sums of the series $\hat{u}(y)$ and $\hat{w}(y)$.

Consider now some horizontal line in the inner region; that is, let
$\text{Im} \,y<0$ be fixed and $\text{Re} \,y$ vary from $-\infty$
to $\infty$.
If the integration contour $\gamma$ is chosen to lie
in the first quadrant of the complex $p$ plane,  the functions $u(y)$ and $w(y)$ generated
by eqs \eqref{laplace} will tend to zero as $\Real y\to +\infty$
along this line.  Similarly,
if it is chosen to lie in the second quadrant, they will tend
to zero as $\Real y\to -\infty$.  Suppose there were no singularities 
between two such contours: then the one could be continously 
deformed to the other
without any singularity crossings; i.e.\ they would 
generate the same solution which, 
therefore, would decay to zero at
both infinities. [That is, the oscillatory tails in
\eqref{expexpansion} would have zero amplitudes, $K_n=0$.]
  In general, however, $U(p)$ and $W(p)$ have singularities
both on and away from the imaginary axis.  In order to minimise the
number of singularities to be crossed in the deformation
of one contour to the other, we choose the contours to lie above
all singularities with nonzero real part.
 (That this is possible,
is shown in appendix \ref{B}.)
Note also that the imaginary part of the singularity
grows faster than its real part and
hence  $\arg p$ should tend to $\pi/2$ as $|p| \to
\infty$ along $\gamma$; this was precisely our choice for
the direction of the contours $\gamma$ in the beginning
of section \ref{BLT}.

\begin{figure}[btp]
\includegraphics{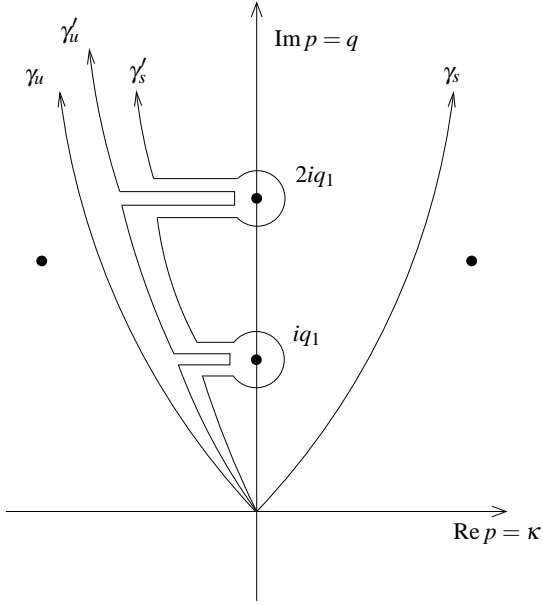}
\caption{\label{integration-contours}The integration contours
$\gamma_s$ and $\gamma_u$ used to generate the solutions
$(u_s, w_s)$ and $(u_u, w_u)$ respectively via eqs \eqref{laplace}.
The dots are singularities of $W(p)$.  Shown  is the
situation where the linear dispersion relation \eqref{leading-order-disp}
has only one real root, $q_1$.}
\end{figure}

Let $\gamma_s$ and $\gamma_u$ be two contours chosen in this
way, with
$\gamma_s$ lying in the first quadrant and $\gamma_u$ in the
second quadrant.
The solutions $u(y), w(y)$ generated by eqs \eqref{laplace} with the contour
$\gamma_s$ will
tend to zero as $\Real y \to \infty$
(with $\text{Im} \, y <0$ fixed). Hence they can be identified with solutions 
$u_s, w_s$ obtained by the continuation of the outer solution  $\psi_s$ 
 which has the
same asymptotic behaviour. 
Similarly, the solutions generated by eqs \eqref{laplace} with the contour $\gamma_u$ 
coincide with solutions $u_u$, $w_u$ -- like $u_u$, $w_u$, the solutions 
generated by eqs \eqref{laplace}   tend to zero 
as $\Real y \to -\infty$ (with fixed $\Imag y <0$).

 Consider, first, solutions $w_s$ and
$w_u$.
Since the contours $\gamma_s$ and $\gamma_u$
 are separated by singularities of  $W(p)$ on
the positive imaginary axis, they
cannot be
continuously deformed to each other
without singularity crossings and so
the solution $w_s$ does not coincide with $w_u$,
unless the residue at the singularity happens to be zero.
If we deform $\gamma_s$ to $\gamma_s'$ and $\gamma_u$ to
$\gamma_u'$ as shown in fig.\,\ref{integration-contours}, without
crossing any singularities, then the only difference between
the two contours is that $\gamma_s'$ encircles the singularities,
whereas $\gamma_u'$ does not. Therefore, the difference $w_s-w_u$
can be deduced exclusively from the leading-order behaviour of  $W(p)$
near its singularities.
There can be two contributions to this difference:
the first arises from integrating around the poles and
is a sum of residues, while the second arises if the singularity is a
branch point.

To find the singularity structure of the function $W(p)$, we equate
\begin{equation}
w_s-w_u= \int_{\gamma_s'} W(p) e^{-py} dp- \int_{\gamma_u'} W(p) e^{-py} dp
\label{wswu}
\end{equation}
to the expansion \eqref{expexpansionb}. The
first term in \eqref{expexpansionb}, $e^{-i q_ny}$, arises
from the integration of a term $(2 \pi i)^{-1} (p-i q_n)^{-1}$ in
$W(p)$. For such a term, the difference of the two integrals in
\eqref{wswu} reduces to an integral around a circle
centered on the point $p=iq_n$:
\[
\frac{1}{2 \pi i} \oint \frac{1}{p-iq_n} e^{-py} dp=
\text{res} \left\{ \frac{e^{-py}}{p-iq_n}, iq_n \right\}.
\]
The term $y^{-m}e^{-iqy_n}$ in \eqref{expexpansionb}, with $m=1,2,\ldots$,
arises from the integration of a term
\[
 \frac{1}{2 \pi i} \frac{(p-iq_n)^{m-1}}{(m-1)!} \ln (p-iq_n)
\]
in $W(p)$. This time, $p=iq_n$ is a branch point. After going around
this point along the circular part of $\gamma_s'$, the logarithm
increases by $2 \pi i$ and the difference between the integrals in
\eqref{wswu} is given by
\begin{align*}
&\frac{1}{(m-1)!} \int_C (p-iq_n)^{m-1} e^{-py} dp \\
&\quad = \frac{e^{-i q_n y}}{(m-1)!} \int_0^\infty z^{m-1} e^{-zy}dz,
\end{align*}
where $C$ is the part of $\gamma_s'$ extending from
$p=iq_n$ to infinity. This equals exactly
$y^{-m} e^{-i q_n y}$.

Thus, in order to generate the full series \eqref{expexpansionb}
we must have
\begin{multline}
\label{Wsingularities}
W(p) =
\frac{1}{2 \pi i} \sum_n K_n
\Bigg[
\frac{1}{p-iq_n}
+ \sum_{m=1}^\infty \frac{c_m^{(n)}}{(m-1)!}\\
\times (p-iq_n)^{m-1} \ln(p-iq_n) \Bigg]
+ W_{\rm reg}(p),
\end{multline}
where $W_{\rm reg}$ denotes the part of $W$ which is regular
at $p=iq_n$, $n=1,2,\ldots$.
By the same process, matching $u_s-u_u$ to $\delta u$ in
\eqref{expexpansiona} yields
\begin{multline}
\label{Usingularities}
U(p) =
\frac{1}{2 \pi i} \sum_n K_n
\sum_{m=1}^\infty \frac{d_m^{(n)}}{(m-1)!}\\
\times (p-iq_n)^{m-1} \ln(p-iq_n)
+ U_{\text{reg}}(p).
\end{multline}

The solution to eqs \eqref{integral-equations} is nonunique;
for instance, if $\{ U(p), W(p) \}$ is a solution,
then so is $\{ e^{py_0+ \zeta_0} U(p), e^{py_0 - \zeta_0}W(p) \}$
with any complex $y_0$ and $\zeta_0$. Also,
if $\{ U(p), W(p) \}$ is a solution, $\{ W(-p), U(-p) \}$ is
another one. We will impose the constraint
\begin{equation}
U(p)= W(-p);
\label{constraint}
\end{equation}
this constraint is obviously compatible with
eqs \eqref{integral-equations}.
It is not difficult to see that the reduction \eqref{constraint}
singles out a unique solution of eqs \eqref{integral-equations}.
The motivation for imposing the constraint \eqref{constraint}
comes from the symmetry $\hat{u}(-y)=-\hat{w}(y)$ of the power-series
solution of eq.\,\eqref{inner-equations}. Using this symmetry in
eq.\,\eqref{uywy}, we get $\rho_\ell= (-1)^\ell \nu_\ell$ and
then eq.\,\eqref{UpWp} implies \eqref{constraint}.

In view of \eqref{constraint}, the singularities of $U(p)$ in the upper
half-plane are singularities of $W(p)$ in the lower half-plane,
which fall within $W_{\text{reg}}(p)$, and vice versa.
Thus we have, finally,
\begin{multline}
\label{allWsingularities}
W(p) =
\frac{1}{2 \pi i} \sum_n
 \frac{K_n }{p-iq_n}
+ \frac{1}{2 \pi i} \sum_n K_n
\sum_{m=1}^\infty \frac{1}{(m-1)!} \\ 
\shoveleft{\times\Big[ c_m^{(n)}(p-iq_n)^{m-1} \ln(p-iq_n)}\\
-(-1)^m d_m^{(n)}
(p+iq_n)^{m-1} \ln(p+iq_n) \Big]
+ \tilde{W}_{\text{reg}}(p),
\end{multline}
where $\tilde{W}_{\text{reg}}(p)$ is regular at $p = \pm iq_n$.
We also mention an equivalent representation for
\eqref{allWsingularities} which turns out to be
computationally advantageous:
\begin{multline}
\label{allWsingularities_int}
W(p) =
 \frac{1}{2 \pi i} \sum_n K_n
\sum_{m=0}^\infty
  D^{-m} \left[\frac{c_m^{(n)}}{p-iq_n} -
 \frac{ (-1)^m d_m^{(n)}}{p+iq_n}
 \right] \\
+ \check{W}_{\text{reg}}(p).
\end{multline}
Here $D^{-1}$ is an integral map:
\[
D^{-1} f(p) \equiv  \int_0^p f(p_1) dp_1;
\]
the notation $D^{-m} f(p)$ should be understood as
\[
D^{-m} f(p) \equiv  \int_0^p \! dp_1 \int_0^{p_1} \! dp_2 \int_0^{p_2}
 \! dp_3 \cdots \int_0^{p_{m-1}} \! dp_m f(p_m).
\]
We have also introduced $c_0^{(n)} = 1$ and
$d_0^{(n)} = 0$ for economy of notation.
The only difference between eqs \eqref{allWsingularities} and
\eqref{allWsingularities_int} is that the double-sum term on the right-hand
side of \eqref{allWsingularities_int} includes some terms
which are regular at $p=\pm iq_n$, whereas in 
eq.\,\eqref{allWsingularities}, all regular terms are contained in
$\tilde{W}_{\text{reg}}(p)$.

The residues $K_n$ at the poles of $W(p)$ are known
as the Stokes constants.  The
leading-order Stokes constant $K_1$
can be related to the behaviour of the
coefficients in the power-series expansion of $W(p)$.
Indeed, the coefficients in the power series \eqref{UpWp} satisfy
\begin{equation}
\label{nudef}
\nu_{\ell} \longrightarrow K_1 \sum_{m=0}^{\ell}\frac{c^{(1)}_m +
(-1)^{\ell}d^{(1)}_m}{2\pi q_1
(i q_1)^{\ell-m}}\frac{(\ell-m)!}{\ell!}
\quad \text{as } \ell \to \infty.
\end{equation}
This is obtained by expanding the
singular part of the expression \eqref{allWsingularities_int}
in  powers of $p$. (Coefficients of the regular part become
negligible in the limit $\ell\to\infty$ compared to those of the
singular part.)  Note that we have ignored singularities
with nonzero real part and singularities on the imaginary axis
other than at $p = \pm iq_1$. The reason is that all these
singularities are further away from the origin than the points
$\pm iq_1$
(in particular all complex singularities are separated
from the origin by a distance greater than $2 \pi$),
and their contribution to $\nu_{\ell}$ becomes
vanishingly small as $\ell \to \infty$.
We have also neglected singularities at the `combination points'
because of their exponentially small residues.
The coefficients $\nu_{\ell}$ can be calculated numerically; once they
are known, it follows from \eqref{nudef} that
\begin{equation}
\label{limit}
K_1 = 2\pi q_1\lim_{\ell\to\infty}(iq_1)^{\ell}\nu_{\ell}
\end{equation}
(where we have recalled that $c_0^{(1)} = 1$ and $d_0^{(1)} = 0$).

We now turn to the numerical calculation of the coefficients $\nu_{\ell}$.

\subsection{Recurrence relation}

To make our forthcoming
numerical procedure more robust,
we normalise  the coefficients in the
power series
 \eqref{UpWp}  by writing
\begin{equation}
\nu_\ell = -i\frac{\delta_\ell}{(iq_1)^\ell}.
\label{delta}
\end{equation}
Substituting the expansions \eqref{UpWp} with eq.\,\eqref{delta}
as well as the constraint \eqref{constraint} into
either of equations
\eqref{integral-equations} and equating coefficients of
like powers of $p$, yields the following
recurrence relation for the numbers $\delta_n$
($n \ge 0$):
\begin{multline}
\sum_{m=0}^n \frac{q_1^m\delta_{n-m}}{(m+2)!}{\cal R}_m
= \frac{1}{(n+2)(n+1)}\\ 
\shoveleft{\times \sum_{m=0}^n\left[\delta_{n-m}
+\mu\sum_{j=2}^{n-m}\frac{q_1^j\delta_{n-m-j}}{j!}
 {\cal R}_j \right]} \\
\times\left(\sum_{j=0}^m (-1)^{m-j}\delta_{m-j}
\delta_j\frac{j!(m-j)!}{m!}\right)\frac{m!(n-m)!}{n!}.
\label{recu}
\end{multline}
Here
\[
{\cal R}_m= \Bigg\{
\begin{array}{ll}
(-1)^{\frac{m}{2}}\cos k, &  \text{for } m \text{ even}, \\
(-1)^{\frac{m-1}{2}}\sin k,  & \text{for } m \text{ odd}.
\end{array}
\]
Solving eq.\,\eqref{recu} with $n = 0$ gives $\delta_0 = \sqrt{1-c^2}$.
Thereafter it can be solved for each member of the sequence
$\{ \delta_n \}$  in terms of the preceding ones, and thus each $\delta_n$
can be calculated in turn.
[Since all the coefficients in the recurrence relation \eqref{recu}
are real, the sequence $\{\delta_n\}$ turns out to be a sequence
of real numbers.]

\begin{figure}[tbp]
\includegraphics{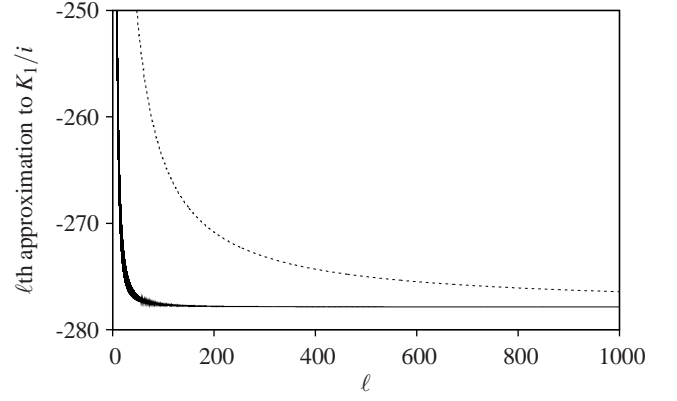}
\caption{\label{series}Convergence of the
sequence on the right-hand side of \eqref{limmo} (dashed line) and
the `accelerated' sequence defined by
the right-hand side of  \eqref{betterlimit}
(solid line).
Shown are the $\ell$th approximations to the Stokes constant $K_1$
[the $\ell$th members of the sequences \eqref{limmo}
and  \eqref{betterlimit}]
 divided by $i$ to get a real value.
In this plot, $\mu = 0$ and $k = 0.5$.
}
\end{figure}

Once the sequence $\{ \delta_n \}$ has been generated,
expression \eqref{limit} can be
used to calculate the Stokes constant $K_1$:
\begin{equation}
K_1 = -2 \pi i q_1  \delta_\ell,
\label{limmo}
\end{equation}
for sufficiently large $\ell$.
Unfortunately, the convergence of the sequence $\{\delta_n\}$
  is slow and thus
the above procedure is computationally expensive.
The convergence can be accelerated by
 expanding  eq.\,\eqref{nudef} in powers of
 small $1/\ell$:
\[
\delta_{\ell} = \frac{iK_1}{2\pi q_1}\!\!\left[1 +iq_1 \frac{c_1^{(1)}}{\ell}
- q_1^2\frac{c_2^{(1)}+(-1)^\ell d_2^{(1)}}{\ell^2} +
\order{\frac{1}{\ell^3}}\right],
\]
whence
\begin{multline}
K_1 = -2\pi i q_1 \delta_{\ell} \Bigg[1 -iq_1 \frac{c_1^{(1)}}{\ell} \\
+ q_1^2\frac{ c_2^{(1)}-(c_1^{(1)})^2 +(-1)^\ell d_2^{(1)}  }{\ell^2}
+ \order{\frac{1}{\ell^3}}\Bigg]. \label{K1}
\end{multline}
According to eq.\,\eqref{K1}, eq.\,\eqref{limmo}  gives $K_1$ with
a relatively large error of order $1/\ell$.
On the other hand, a two-term approximation
\begin{equation}
\label{betterlimit}
K_1 = -2\pi i q_1 \delta_{\ell}\left(1 -iq_1 \frac{c_1^{(1)}}{\ell} \right)
\end{equation}
is correct to  $\mathcal{O} (1/\ell^2)$.  More precisely, the
relative error associated with the answer \eqref{betterlimit} is given by
\begin{equation}
\frac{{\mathcal E}}{K_1} =q_1^2 \frac{c_2^{(1)}-(c_1^{(1)})^2 +(-1)^\ell
d_2^{(1)}}{\ell^2}.
\label{relati}
\end{equation}
In our calculations, we set $\mathcal{E}/K_1=10^{-5}$.
Since $c_1^{(1)}$, $c_2^{(1)}$
and $d_2^{(1)}$ are known constants [given by
\eqref{c1c2} and \eqref{d1n}], eq.\,\eqref{relati}  tells us what
$\ell$ we should take -- i.e.\ how many members of the sequence $\{\delta_n\}$
we should calculate in order to achieve the set accuracy.
Figure \ref{series} illustrates the
convergence of the
approximate values of $K_1$ calculated using \eqref{limmo} and
\eqref{betterlimit} as $\ell$ is increased.  Note the drastic
acceleration of convergence
in the latter case.

\begin{figure}[tbp]
\includegraphics{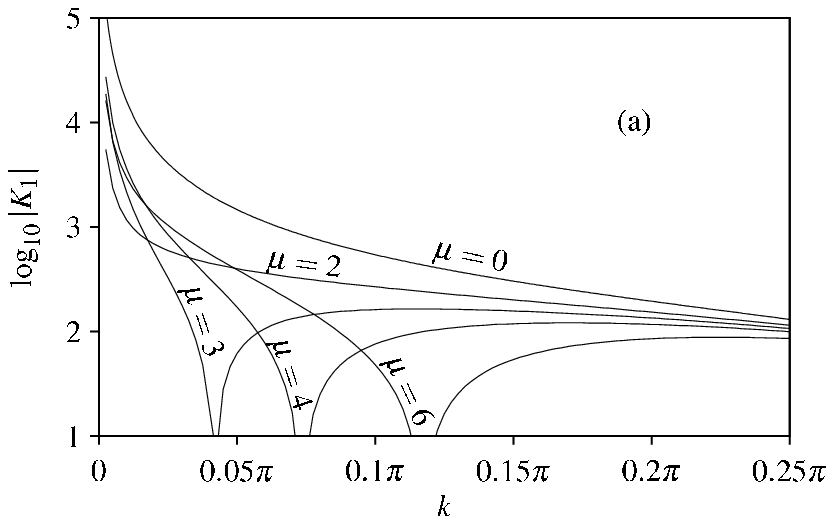}

\medskip
\includegraphics{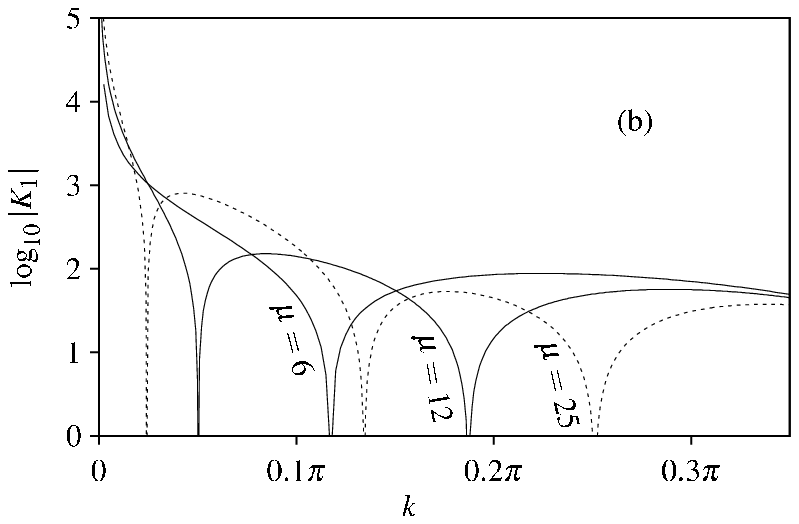}
\caption{\label{stokes}The Stokes constant $K_1$ for various
values of $\mu$.
Note the logarithmic scale on the vertical axis.
The
 downward spikes extend all the way
 to $-\infty$; hence each spike corresponds to a zero crossing.
 Each panel shows only a portion of the full range
 $0 \leq k \leq \pi/2$; there
 are no additional zero crossings in the part which is not shown.
 }
\end{figure}

Figure \ref{stokes}(a) shows the calculated Stokes constant as a
function of $k$ for various values of
the saturation parameter $\mu$.
First of all, $K_1(k)$
 does not have any zeros
in the case of the cubic nonlinearity ($\mu=0$).
This means that solitons of the
cubic one-site discrete NLS equation [eq.\,\eqref{DNLS} with $\mu=0$]
cannot propagate without losing energy to radiation.
For $\mu = 3$ the Stokes constant
 does have a zero, but at a value of $k$ smaller than $k^{(1)}_{\rm max}$
(where $k^{(1)}_{\rm max}\approx0.22=0.07\pi$).
Since higher radiation modes do exist in this range of $k$,
there will still be radiation from the soliton -- unless the
`higher' Stokes constants $K_2(k)$, $K_3(k)$,\ldots,  happen to be zero
at the same value of $k$.
Finally,
for $\mu = 4$ the zero is seen to have moved just above $k^{(1)}_{\rm max}$ and
for $\mu = 6$ it has an even higher value.
There are no $\mathcal{Q}_2, \mathcal{Q}_3,\ldots$ radiations
for these $k$; hence the zeros of $K_1(k)$ define
the carrier wavenumbers at which the soliton `slides' --
i.e.\ travels without emitting any radiation. Equation \eqref{om_and_v_via_eps}
then gives the corresponding sliding velocities, for each $\epsilon$.

Figure \ref{stokes}(b) shows the  Stokes constant $K_1(k)$
for higher values of the
parameter $\mu$.  For $\mu = 12$ a second zero of the
Stokes constant has appeared while for $\mu = 25$,
the function $K_1(k)$ has three zeros.
As $\mu$ is increased, the existing zeros move to
larger values of $k$  while new
ones emerge  at the origin of the $k$ axis.

\subsection{Radiation waves}

For not very large $|X|$, the solution $\psi_s$ is close to the localised pulse
found by means of the perturbation expansion
in section \ref{section2}.  As $X \to +\infty$, it tends
to zero, by definition, while the $X \to -\infty$ asymptotic
behaviour is found from  $\psi_s = \psi_u + \Psi$.
 Here the solution $\psi_u$ decays to zero as $X \to
-\infty$ and hence
$\psi_s$ approaches the oscillatory waveform
 $\Psi$
given by eqs \eqref{radansatz}, \eqref{f1g2}, and \eqref{symb}:
\begin{multline}
\label{final-solution}
\psi_s(X) \to \sum_n K_n^*e^{-\pi \mathcal{Q}_n/2\epsilon} \\
\times\left[1 +\epsilon
\frac{4  i\cos k\tan X}{2\sin(k+\mathcal{Q}_n)-v} +
\order{\epsilon^2}\right]e^{i\mathcal{Q}_n X/\epsilon}
\end{multline}
 as $X\to -\infty$.
Equation \eqref{final-solution} describes a radiation background over which
 the soliton is superimposed.
 As we have explained, we can ignore all but the first term in the
sum.

To determine whether the radiation is emitted by the soliton
or being fed into it from outside sources,
 we
consider a harmonic solution $\psi =  e^{i\mathcal{Q}X/\epsilon-i\Omega t}$
of the linearised eq.\,\eqref{timedepeqn}; the
corresponding dispersion relation is
\begin{equation*}
\Omega(\mathcal{Q}) = -2\cos(k+\mathcal{Q}) + \omega - \mathcal{Q}v.
\end{equation*}
The  radiation background $\Psi$ consists of
harmonics with
$\Omega = 0$ and $\mathcal{Q} = \mathcal{Q}_n$
where  $\mathcal{Q}_n$ are roots of eq.\,\eqref{disprel}.
The group velocities of these harmonic waves are given by
\begin{equation}
\Omega'(\mathcal{Q}_n) = 2\sin(k+\mathcal{Q}_n)-v.
\label{vgr}
\end{equation}
The $\mathcal{Q}_n$'s are zeros of the function $\Omega(\mathcal{Q})$
and the group velocities are the slopes of this function at its zeros;
thus the group velocities
$\Omega'(\mathcal{Q}_1)$, $\Omega'(\mathcal{Q}_2)$,\ldots,
 have alternating signs.  The first one,
which is the only one that concerns us in this work,
 must be negative.  Indeed, the value
\[
\Omega(0) = -2\cos k+\omega = 2\cos k (\cosh\epsilon-1)
\]
is positive, and hence the
  slope of the
function $\Omega(\mathcal{Q})$ as it crosses the $\mathcal{Q}$ axis
at $\mathcal{Q}_1>0$  is negative.
Therefore, the first radiation
mode, extending to $-\infty$,  carries energy {\it away\/} from the
soliton.

The even-numbered radiation modes (where present) in our
asymptotic solution \eqref{final-solution} have positive group
velocities
and hence describe the flux of energy fed into
 the system at the left infinity.
A more interesting situation is obviously the one with
no incoming radiation; the corresponding solution is obtained by
subtracting off the required multiple of the solution of the
linearised equation, e.g.\ $e^{i\mathcal{Q}_2X/\epsilon}$.  One would then
have a pulse leaving the odd modes in its wake and sending even
modes ahead of it.

If the first Stokes constant $K_1(k)$ has a zero at
some $k=k_1$ while $\mathcal{Q}_1$ is the only
radiation mode available (as happens in our
saturable model with $\mu$ greater than
approximately $4$), then according to
eq.\,\eqref{final-solution}, the radiation from
the soliton with the carrier-wave wavenumber $k_1$ is
suppressed completely.

\section{Time evolution of a radiating soliton}
\label{section4}

\subsection{Amplitude-wavenumber dynamical system}

To find the radiation-induced evolution of the
travelling soliton, we use conserved quantities of the advance-delay
equation associated with eq.\,\eqref{DNLS}.
In the reference frame moving at the
soliton velocity $v$ this equation reads
\begin{equation}
\label{timedepeqn2}
i\varphi_t +\varphi(x+1,t)  + \varphi(x-1,t) -iv \varphi_x
+\frac{2|\varphi|^2\varphi}{1+\mu|\varphi|^2} = 0.
\end{equation}
The discrete variable $\phi_n(t)$ in eq.\,\eqref{DNLS} is related to
the value of the continuous variable $\varphi(x,t)$ at the
point $x=n-vt$:  $\phi_n(t) = \varphi(n-vt,t)$.
For future use, we also mention the relation between $\varphi(x,t)$ and
 the corresponding solution of eq.\,\eqref{timedepeqn}:
 \begin{equation}
 \varphi(x,t)= \psi(X,t) e^{ik(x+vt)+ i \omega t}.
 \label{corresp}
 \end{equation}

We first consider the number of particles integral:
\[ N = \int_{a}^{b}|\varphi|^2dx. \]
Multiplying eq.\,\eqref{timedepeqn2}
by $\conj{\varphi}$, subtracting the complex
conjugate and integrating yields the rate of change of the integral $N$:
\begin{align}
\label{mainN}
 i\frac{dN}{dt} &= \int_{a-1}^{a}  (\varphi^+\conj{\varphi} - c.c.) dx \notag\\
&\quad + \int_b^{b+1} (\varphi^-\conj{\varphi} - c.c.) dx
+  iv|\varphi|^2 \Big|_{a}^b.
\end{align}
In eq.\,\eqref{mainN}, $\varphi^\pm \equiv \varphi(x \pm 1,t)$
and    $c.c.$ stands for the complex conjugate of the
immediately preceding term.
The  integration limits $a<0$ and $b>0$ are
 assumed to be large ($|a|,b \gg \epsilon^{-1}$)
but finite; for example one can take $a,b= {\cal O}(\epsilon^{-2})$.

Consider the soliton moving with a positive velocity
and leaving radiation in its wake. This
configuration is described by the
solution $\psi_s$ of eq.\,\eqref{timedepeqn}; the corresponding
solution of eq.\,\eqref{timedepeqn2} has the asymptotic behaviour
 $\varphi_s \to 0$ as $x \to +\infty$.
Substituting the leading-order expression  \eqref{stable-minus-unstable-outer}
for the soliton's radiation tail into eq.\,\eqref{corresp} yields
the  asymptotic behaviour at the other infinity:
\begin{equation}
\varphi_s(x,t) \to \sum_n A_n[1+\order{\epsilon}]e^{i(k+{\cal Q}_n)x+i(\omega+kv)t}
\label{asympie}
\end{equation}
as $x \to -\infty$.
Since, as we have explained above, the $\mathcal{Q}_1$ radiation is dominant, it
is sufficient to keep only the $n=1$ term in \eqref{asympie}.

Substituting \eqref{asympie} into \eqref{mainN} and evaluating the
integral over the region $(a-1,a)$ in the soliton's wake,
 we obtain
\begin{equation}
\label{Ndot}
\frac{dN}{dt} =
|K_1|^2e^{-\pi \mathcal{Q}_1/\epsilon}[2\sin(k+\mathcal{Q}_1) - v],
\end{equation}
where we have used eq.\,\eqref{symb}.
Note that $[2\sin(k+\mathcal{Q}_1) - v]$ is the group velocity
of the first radiation
mode, $\Omega'(\mathcal{Q}_1)$, which, as we have established, is negative;
 hence ${\dot N} \leq 0$.

We now turn to the momentum integral,
\[
P =
\frac{i}{2}\int_{a}^b (\conj{\varphi}_x\varphi-\varphi_x\conj{\varphi}) dx.
\]
Multiplying \eqref{timedepeqn2} by $\conj{\varphi}_x$, adding its complex
conjugate and integrating, gives the rate equation
\begin{multline}
\label{mainP}
\frac{dP}{dt} =
\int_{a-1}^{a} (\varphi^+\conj{\varphi}_x + c.c.) dx
-\int_b^{b+1} (\varphi^-\conj{\varphi}_x + c.c.) dx \\
+ \frac{i}{2}  (\varphi \conj{\varphi_t}
-\conj{\varphi} \varphi_t) \Big|_{a}^b
-   (\varphi^+)^* \varphi \Big|_{a-1}^b
- (\varphi^-)^* \varphi \Big|_{a}^{b+1} \\
-\left[\frac{2}{\mu}|\varphi|^2
-\frac{2}{\mu^2}\ln\left(1+\mu|\varphi|^2\right)\right]_{a}^b.
\end{multline}
Evaluating the right-hand side of
\eqref{mainP} similarly to the
way we obtained \eqref{Ndot} and substituting eq.\,\eqref{disprel}
for $\omega$,
produces
\begin{equation}
\label{Pdot}
\frac{dP}{dt} =
|K_1|^2 e^{-\pi \mathcal{Q}_1/\epsilon}(k+\mathcal{Q}_1)[2\sin(k+\mathcal{Q}_1) - v].
\end{equation}

Using the leading-order term of the perturbative solution
\eqref{regularpert} and eq.\,\eqref{corresp},
we can express $N$ and $P$
via $\epsilon$ and $k$:
\begin{eqnarray*}
N  =2\epsilon\cos k + \order{\epsilon^2}, \\
P = 2k \epsilon\cos k + \order{\epsilon^2}.
\end{eqnarray*}
In calculating $N$ and $P$
we had to integrate from  $X =\epsilon a$ to $X=\epsilon b$.
  Since the integrands decay exponentially, and since
  $a<0$ and $b>0$ were assumed to be much larger than $\epsilon^{-1}$
  in absolute value, it was legitimate to replace these
  limits with $-\infty$ and $\infty$, respectively.
  The error introduced in this way is exponentially small in $\epsilon$.

Taking time derivatives of $N$ and $P$ above
and discarding $\epsilon^1$ corrections to $\epsilon^0$ terms,
 we deduce that
\begin{align*}
{\dot \epsilon}   & = \frac{\dot{N}+\tan k(\dot{P}-k\dot{N})}{2\cos k},
\\
{\dot k} & =\frac{\dot{P}-k\dot{N}}{2\epsilon\cos k}.
\end{align*}
Finally,
 substituting for $\dot{N}$ and $\dot{P}$ from eqs \eqref{Ndot} and \eqref{Pdot},
  we arrive at the dynamical system
\begin{subequations}
\label{paramev}
\begin{align}
\dot{\epsilon} &= |K_1(k)|^2e^{-\pi \mathcal{Q}_1/\epsilon} \,
\Omega'(\mathcal{Q}_1) \,
\frac{1+\mathcal{Q}_1\tan k}{2\cos k},
\label{param_top} \\
\dot{k} &= |K_1(k)|^2e^{-\pi \mathcal{Q}_1/\epsilon} \,
\Omega'(\mathcal{Q}_1) \,
\frac{\mathcal{Q}_1}{2\epsilon\cos k}, \label{param_bottom}
\end{align}
\end{subequations}
where the group
velocity $\Omega'(\mathcal{Q}_1)=2 \sin(k+\mathcal{Q}_1)-v$ with $v=2 (\sinh \epsilon / \epsilon)
\sin k$,
and $\mathcal{Q}_1=\mathcal{Q}_1(\epsilon,k)$ is the 
smallest root of eq.\,\eqref{nicedisprel}.

\begin{figure}[b]
\includegraphics{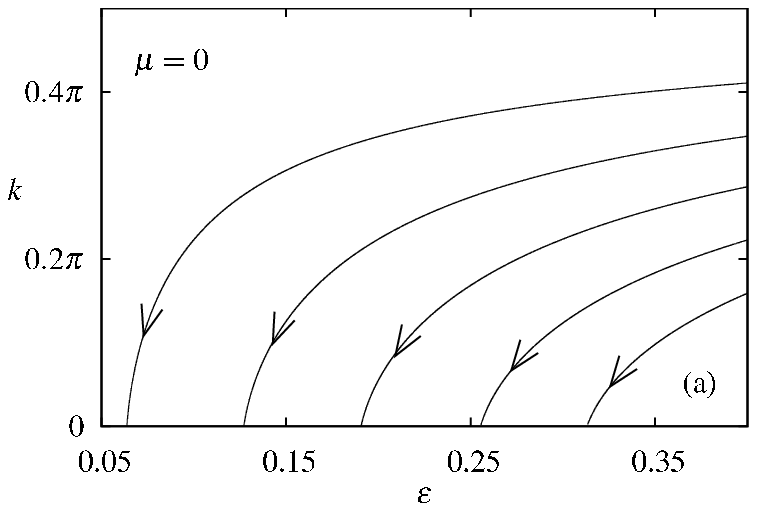}

\medskip
\includegraphics{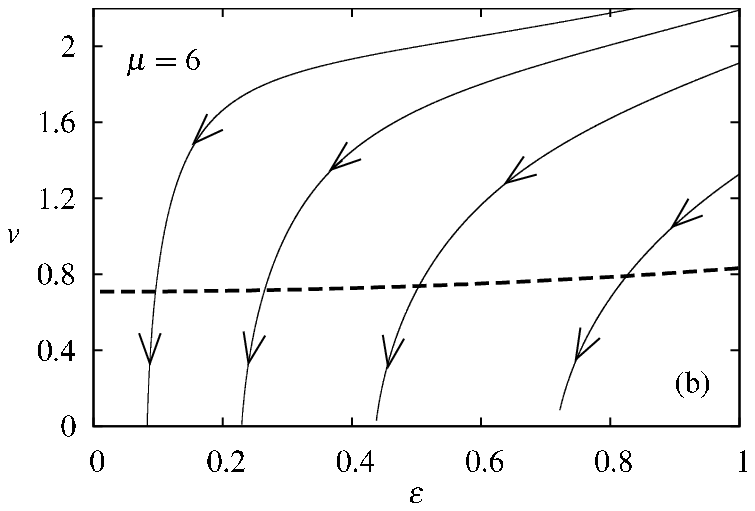}
\caption{\label{phaseplane} The phase portrait of the system \eqref{paramev}
in the case (a) where the Stokes constant $K_1(k)$ does not have zeros, and (b) 
where $K_1(k)$ has one zero. In (b), the dashed line is the line of nonisolated 
fixed points $k=k_1$. Note that in (b), the phase portrait has been replotted on
the $(\epsilon, v)$ plane; hence the dashed line gives the value of the sliding 
velocity for each value of $\epsilon$. In (a), $\mu=0$; in (b), $\mu=6$.}
\end{figure}

\subsection{Soliton's deceleration and sliding velocities}

The vector field \eqref{paramev} is defined for $k \geq k^{(1)}_{\rm min}$,
where  $k^{(1)}_{\rm min}$ is the value of $k$
for which the roots $\mathcal{Q}_1$ and $\mathcal{Q}_2$ merge in
fig.\,\ref{nicedisprelfig} (i.e.\ the smallest value of $k$ for
which the roots  $\mathcal{Q}_1$ and $\mathcal{Q}_2$
still exist). When $k = k^{(1)}_{\rm min}$, the group velocity $\Omega'(\mathcal{Q}_1)$
 becomes equal to zero; therefore,
 the equation $k=k^{(1)}_{\rm min}(\epsilon)$ defines a line of
(nonisolated) fixed points of the system
\eqref{paramev}.
Assume, first, that
the saturation parameter $\mu$ is such that $K_1(k)$ does not vanish for
any $k$.
 For $k > k^{(1)}_{\rm min}$,
the factor $\Omega'(\mathcal{Q}_1)$ in \eqref{param_bottom} is negative;  since
$\mathcal{Q}_1$ and $\cos k$ are
positive for all $0< k \leq \pi/2$, the derivative
$\dot{k}$ satisfies
$\dot{k} \leq 0$ for all times. Hence, all trajectories are flowing towards the
line $k=k^{(1)}_{\rm min}(\epsilon)$ from above.
It is worth
noting that the derivative
$\dot{\epsilon}$ is also negative, and that the $k$ axis is also
a line of nonisolated fixed points. However, no
trajectories will end there as follows from the equation
\[
\frac{d \epsilon}{dk} = \epsilon \left[ \tan k +
\frac{1}{\mathcal{Q}_1(k,\epsilon)} \right],
\]
which is a consequence of \eqref{param_top} and \eqref{param_bottom}.

Representative trajectories are plotted in fig.\,\ref{phaseplane}(a).
Since $k^{(1)}_{\rm min}(\epsilon) \approx \epsilon^2/4 \pi$
is very small for small $\epsilon$,
the line $k=k^{(1)}_{\rm min}(\epsilon)$ is practically indistinguishable
from the horizontal axis. Therefore, we can assert that, to a good
accuracy,
$k \to 0$ as $t \to \infty$.
This means that the soliton stops moving --
and stops decaying at the same time.

For $k$ smaller than $k^{(1)}_{\rm min}$, the vector field \eqref{paramev} is
undefined and we cannot use it to find out what happens
to the soliton after $k$
has reached $k^{(1)}_{\rm min}$. The reason
for this is that the soliton stops radiating
at the wavenumber $\mathcal{Q}_1$
as $k$ drops below $k^{(1)}_{\rm min}$. In fact our analysis becomes invalid
as soon as $k$ becomes $\mathcal{O}(\epsilon^2)$---i.e.\ even before $k$
 reaches $k^{(1)}_{\rm min}$---because we can no longer disregard the
$n=2$ radiation here.
At the qualitative level it
is obvious, however, that the parameter $k$ should continue
to decay all the way to zero, in a cascade way.
First, the $n=2$ radiation will
become as intense as the $n=1$ mode when $k$ approaches $\epsilon^2/4\pi$.
Subsequently---i.e.\ for $k$ smaller than $\epsilon^2/4\pi$---the $n=3$ harmonic will
replace the radiation with the wavenumbers $\mathcal{Q}_1$ and
$\mathcal{Q}_2$ as a dominant mode.
 The $n=4$ mode will become equally intense
 near $k=k^{(2)}_{\rm min} \approx \epsilon^2/8\pi$; as $k$ drops below $\epsilon^2/8\pi$,
 both  $\mathcal{Q}_3$ and $\mathcal{Q}_4$ will cede to
$\mathcal{Q}_5$, and so on.

If $\mu$ is such that
the Stokes constant $K_1(k)$ vanishes at one or more values of $k$,
the system has one or more lines of nonisolated fixed points, $k=k_i$.
The corresponding values of $v$, $v=v_i(\epsilon)
 \equiv 2 (\sinh\epsilon /\epsilon ) \sin
k_i$, define the {\it sliding\/} velocities of the soliton --
i.e.\ velocities at which the soliton moves without radiative friction.
One such velocity is shown by the dashed line in fig.\,\ref{phaseplane}(b).
The fixed points $(\epsilon, k_i)$
are semistable; for $k$ above
$k_i$, the
flow is towards the line $k=k_i$ but when $k$ is below  $k_i$,
the flow is directed away from this straight line [see fig.\,\ref{phaseplane}(b)].
Since both $\dot{\epsilon}$ and $\dot{k}$ are negative,
the soliton's velocity $v = 2 (\sinh\epsilon /\epsilon ) \sin k$
will generally be decreasing -- until it hits the nearest underlying
sliding velocity $v_i$ and locks on to it.
The ensuing sliding motion will be unstable; a small
perturbation will be sufficient to take the soliton out of the sliding regime
after which it will resume its radiative deceleration.
However, since $\dot{k}$ is proportional to the {\it square\/} of the
Stokes constant and not to $K_1(k)$ itself,  small perturbations
$\delta k$ will be growing linearly, not exponentially, in $t$.
As a result,
the soliton may spend a fairly long time sliding  at the velocity $v_i$.
It is therefore not unreasonable to classify this sliding motion
as {\it metastable\/}.

  We conclude
that the soliton becomes pinned (i.e.\ $k \rightarrow 0$ and so $v
\to 0$) before it has
decayed fully (i.e.\ before the amplitude $\epsilon$ has decreased to zero).
The exponential dependence
of $\dot \epsilon$ and $\dot k$ on $1/\epsilon$ implies that
tall, narrow pulses will be pinned very quickly while short, broad ones
will travel for a very long time before they have slowed appreciably.
Next, if $\mu$ is such that there is one or more sliding velocity
available in the system,
and if the soliton is initially moving faster than some of these,
its deceleration will be interrupted by long periods of undamped
motion at the corresponding sliding velocities.

\begin{figure}
\hspace*{-2.5em}\includegraphics{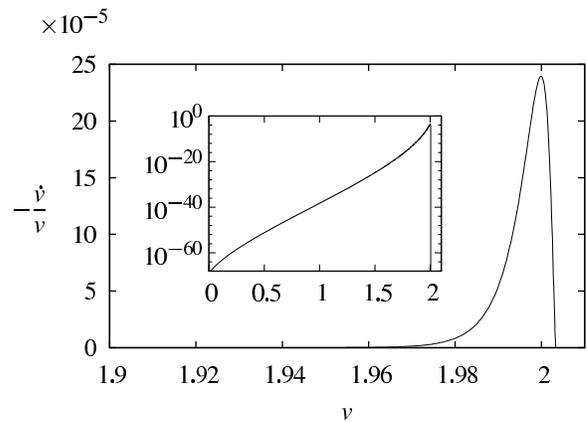}
\caption{\label{decayrate} The soliton's decay rate as a function of its velocity, for fixed $\epsilon=0.1$.
(Here $\mu=0$.) In the inset, the same curve is replotted using a logarithmic scale
on the vertical axis.}
\end{figure}

It is worth reemphasising here that if the 
 amplitude $\epsilon$ is small, then 
even if the soliton is {\it not\/} 
sliding, its deceleration will be so slow that it will spend an exponentially long time travelling
with virtually unchanged amplitude and speed. 
The deceleration rate $-{\dot v}/v$ is shown in fig.\,\ref{decayrate}, as a function of 
the soliton's velocity $v$, for fixed amplitude $\epsilon$. 
Note that the decay rate drops, exponentially, as the velocity is decreased;
this drop is due to the exponential factor $e^{-\pi \mathcal{Q}_1/ \epsilon}$ in eq.\,\eqref{paramev}.
As the velocity $v$
(and hence the wavenumber $k$) decreases, the root $\mathcal{Q}_1(\epsilon,k)$ grows
towards the limit value of approximately $2 \pi$ (see fig.\,\ref{nicedisprelfig}). This 
 variation in $\mathcal{Q}_1$ is amplified by the division by small $\epsilon$
and exponentiation in $e^{-\pi \mathcal{Q}_1/ \epsilon}$.

\section{Concluding remarks}
\label{section5}

\subsection{Summary}

In this paper, we have constructed the moving discrete soliton of the
saturable NLS equation \eqref{DNLS} [and hence \eqref{VK}]
as an asymptotic expansion in powers of its amplitude.
The saturable nonlinearity includes the cubic NLS as a particular case
(for which $\mu=0$). Our perturbation procedure is a variant of the Lindstedt--Poincar\'e
technique where corrections to the parameters of the solution are
calculated along with the calculation of the solution itself.
Although the resulting asymptotic series \eqref{regularpert} for the soliton
is generally not convergent,
the associated expansions for its frequency
and velocity sum up to exact explicit expressions
\eqref{om_and_v_via_eps}.

From the divergence of the asymptotic series \eqref{regularpert}
it follows, in particular,
that the travelling discrete soliton does not decay to zero at least at one of the
two infinities. Instead, the soliton approaches, as
$X \to \infty$ or $X \to -\infty$, an oscillatory resonant background
\eqref{stable-minus-unstable-outer}
where the amplitudes $A_n$ of its constituent harmonic waves
lie beyond all orders of $\epsilon$. For the soliton moving with a
positive velocity and approaching zero as $X \to \infty$, the oscillatory
background at the left infinity represents
the Cherenkov radiation left in the soliton's wake. To
evaluate the amplitudes of the harmonic waves
arising as $X \to -\infty$, we have
continued the radiating soliton
into the complex plane, where it exhibits
singularities. We then matched the asymptotic expansion \eqref{cont12}
of the background
near the lowest singularity on the imaginary $X$ axis to the far-field asymptotic
expansion \eqref{expexpansion} of
the background solution of the `inner' equation -- i.e.\ of the
advance-delay equation `zoomed in' on this singularity. The asymptotic
expansions here are in inverse powers of the zoomed variable, $y$.
The amplitudes of the radiation waves were found to be exponentially
small in $\epsilon$, with the pre-exponential factors (the so-called Stokes
constants) being dependent only on the soliton's carrier-wave wavenumber.
Representing solutions to the inner equation 
  as  Borel sums
of their asymptotic expansions,
the Stokes constants can be related to the expansion coefficients;
we have calculated these coefficients numerically, using algebraic recurrence
relations.

The upshot of the calculation of the leading Stokes constant $K_1$ is that
in the case of the cubic nonlinearity (i.e.\ for $\mu=0$), $K_1(k)$ does not vanish
for any $k$. This means that the cubic discrete soliton cannot `slide' --
i.e.\ cannot move without radiative friction. The saturable
solitons, on the other hand, {\it can\/} slide provided the saturation
parameter $\mu$ is large enough. 
This is because, for a sufficiently large $\mu$, the
Stokes constant $K_1(k)$ is found to have one or more zeros $k_1, k_2, \ldots$.
Since the soliton with wavenumber $k>0.22$ can have no higher-order 
resonances, 
those zeros which satisfy $k_i>0.22$ do define the wavenumbers at which sliding motion
occurs.
For each
value of the soliton's amplitude $\epsilon$, the formula \eqref{om_and_v_via_eps}
then gives the sliding velocities $v(k_i, \epsilon)$.

The calculation of asymptotics beyond all orders is useful not only for
determining the sliding velocities. Knowing the radiation amplitudes has
allowed us to derive a two-dimensional dynamical system
\eqref{paramev} for the soliton's parameters. Trajectories 
of this dynamical system describe the evolution of the soliton 
travelling at a generic speed. The evolution turns out to be 
simple:
If  $\mu$ is such that the Stokes constant
$K_1(k)$ does not vanish anywhere in the region $k> 0.22$, 
the soliton decelerates, although very slowly.
Eventually it becomes pinned to the lattice
with a decreased but finite amplitude.
However, if $\mu$ is such that there are sliding velocities
in the system, and if the soliton starts its motion
with the velocity higher than some of these then, although
it will finally become pinned to the lattice, its 
deceleration will be interrupted by long periods of metastable sliding
motion at these isolated velocities.

\subsection{Concluding remarks}

It is interesting to tie up the above results with our
previous work on exceptional discretisations of the $\phi^4$
model \cite{OPB}. In that case we discovered that for some
exceptional models (in which the stationary soliton possesses an
effective translational symmetry), sliding velocities, at which 
the radiation disappears, do exist.
The fact that all these models involve a
complicated nonlocal discretisation of the nonlinearity
leads one to wonder whether the nonlinearity
{\it has\/} to be discretised nonlocally 
in order for sliding solitons to exist. 
This question is answered%
---negatively---by
 our present work which gives a counterexample of 
 a simple, local and physically motivated
nonlinearity supporting sliding solitons.

A remaining open issue is whether exceptionality of the system is a
prerequisite for the existence of sliding solitary waves. Indeed, we
have yet to encounter a nonexceptional discrete system 
permitting sliding motion. 

We conclude this section by placing our results in the context of
earlier studies of discrete solitons.

Moving solitons in the cubic DNLS equation 
have previously been studied by Ablowitz, Musslimani, and
 Biondini \cite{ablowitz}
(among others), using a numerical technique based on discrete
Fourier transforms
as well as perturbation expansions for small velocities.  As
we have done, these authors suggest that sliding (radiationless)
solitons  may not exist.  They also observe,
with the aid of numerical simulations, that strongly
localised pulses are pinned quickly to the lattice, while
broader ones are more mobile -- results which are in agreement
with ours.

Duncan, Eilbeck, Feddersen and  Wattis   \cite{Feddersen,duncan} have studied the
bifurcation of periodic travelling waves from constant
solutions in the cubic DNLS equation, using numerical
path-following techniques.  They find that the paths terminate
when the soliton's amplitude reaches a certain limit value,
and in the light of our results it seems likely that this 
is the point at which the radiation becomes large enough to
have an effect on the numerics.

Solitary waves which decay to constant values at the spatial infinities,
despite the fact that the generic asymptotic
 behaviour in the underlying model is oscillatory,
are commonly referred to as embedded solitons. An example is given
by the sliding solitons reported in this paper as well as the
sliding kinks of \cite{OPB}; both types of solitary waves propagate without
exciting the resonant background oscillations.
Embedded solitons have been studied for some time in continuous systems
\cite{embedded_review}, but their history in lattice equations is
younger.  Recently, Malomed, Gonz\'alez-P\'erez-Sandi, Fujioka,
Espinosa-Cer\'on and Rodr\'iguez have considered certain lattice equations
with next-to-nearest-neighbour couplings and shown (by means of explicit
solutions) that both stationary
\cite{Malomed_embedded_0} and moving \cite{Malomed_embedded} embedded solitons exist.  Stationary embedded
solitons in discrete waveguide arrays have also been analysed by
Yagasaki, Champneys and Malomed \cite{Malomed_embedded_2}.

Finally, while preparing the revised version of this paper,
we learnt that Melvin, Champneys, Kevrekidis, and Cuevas
have obtained results
very similar to ours. Using a combination of
intuitive arguments
and numerical computations, they have found 
sliding solitons for
certain values of the
parameters of a saturable DNLS model \cite{melvin}.

\begin{acknowledgments}
The authors would like to thank Dmitry Pelinovsky and Alexander Tovbis
for helpful discussions relating to this work.
O.O. was supported by the National Research Foundation of South Africa
and the University of Cape Town.
I.B. was supported by the NRF under
grant 2053723 and by the URC of UCT.

\end{acknowledgments}

\appendix
\section{Convolution theorem for the Laplace transform in the complex plane}
\label{A}

This appendix deals with the convolution property 
of the modified Laplace
transform of the form \eqref{laplace}, where the integration is
along an infinite contour in the complex plane rather than the
positive real axis. In the context of asymptotics beyond
all orders, this transform was pioneered by Grimshaw and Joshi
\cite{GrimshawJoshi}. Since no proof of the convolution result \eqref{conv_thm} 
is available in literature, and since it requires a nontrivial
property (concavity) of the integration contour, we produce such
a proof here.

We wish to show that
\begin{align}
\label{rtp}
&\int_{\gamma}e^{-pz}F(p)dp \int_{\gamma}e^{-p'z}G(p')dp' \nonumber\\
= &\int_{\gamma}e^{-pz}
\left[ \int_0^pF(p_1)G(p-p_1)dp_1  \right] dp,
\end{align}
where $\int_0^p$ stands for an integral along the curve $\gamma$
from the origin to the point $p$ on that curve.
We assume that the curve
$\gamma$ extends from the origin to infinity on the complex plane; lies in 
its first quadrant ($\Real p>0$, $\Imag p>0$);
is described as a graph of a single-valued function $\Imag p= f( \Real p)$
(i.e.\ never turns back on itself),
 and is concave-up everywhere:
\[
\frac{d^2 f}{d(\Real p)^2} > 0 \quad \text{for} \ p \ \text{on} \, \gamma.
\]
We also assume that the function $G(p)$ is analytic in the region between the contour
$\gamma$ and the imaginary axis.

\begin{figure}
\includegraphics{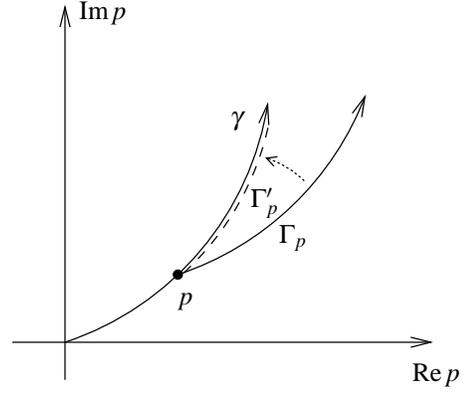}
\caption{\label{conv-paths}Deformation of the integration contour $\Gamma_p$ to $\Gamma_p'$.}
\end{figure}

We begin by writing the left-hand side of \eqref{rtp} as
\begin{align}
  &\int_{\gamma}\int_{\gamma}e^{-(p+p')z}F(p)G(p')dp'dp 
\nonumber \\
\label{undeformed-integral}
= &\int_{\gamma}F(p) \left[ \int_{\Gamma_p}e^{-rz}G(r-p)dr \right] dp,
\end{align}
where $r = p' + p$.  The curve $\Gamma_p$ is the path traced out
by $r$ as $p'$ traces out the curve $\gamma$, for a given
$p$ (which also lies on $\gamma$) -- this is depicted in 
fig.\ \ref{conv-paths}.  The path $\Gamma_p$ is the same
as the path $\gamma$ but translated from the origin to
the point $p$.  

For each given $p$ we deform the path $\Gamma_p$ so that
it now lies on $\gamma$, still starting at the point $p$.
We call this deformed path $\Gamma_p'$.  The point $r-p = p'$, which
lay on the path $\gamma$ before the
deformation, 
will now move inside the region bounded by $\gamma$ (i.e.\ the region to the left of $\gamma$),
since all points on $\Gamma_p'$ lie inside the region bounded by $\Gamma_p$. 
 This follows from the fact that the path $\Gamma_p$ is concave-up.
The point $r-p$ will, however,
stay to the right of the imaginary axis, since after the deformation
all points $r$ on $\Gamma_p'$ lie to the right of the point $p$.
This follows from the fact that the curve
$\gamma$ never turns back on itself.
Since $G(p')$ is analytic for all $p'$ between the imaginary axis and $\gamma$,
the value of the integral \eqref{undeformed-integral}  will not be affected by
the deformation.
[In the particular case of the functions $U(p)$ and $W(p)$ 
considered in section \ref{BLT}, we have deliberately chosen the contours of integration 
so that the analyticity condition is satisfied.]

After the deformation, both integrals in \eqref{undeformed-integral}
follow the path $\gamma$, with the inner one starting at the point $p$.
By parameterising the path $\gamma$ it is now straightforward
to change the order of integration, and end up with the desired
result \eqref{rtp}.

The argument holds, with the obvious modifications, for paths
in the second quadrant.

\section{Proof that singularities do not accumulate to the imaginary axis}
\label{B}

The roots of eq.\,\eqref{singularities} 
with the top and bottom signs give singularities of
$U(p)$ and $W(p)$, respectively.  Our aim
here
is to show that  the integration contours $\gamma_u$ and $\gamma_s$ 
in section \ref{BLT} can be
chosen to lie above all the complex singularities -- i.e., that
singularities of
$U(p)$ and $W(p)$ do not accumulate to the imaginary axis.  

The roots of
eq.\,\eqref{singularities} are zeros of the functions $F(\pm p)$, where
\begin{equation}
F(p) \equiv \cosh p - 1 - i\tan k(\sinh p - p).
\end{equation}
The imaginary zeros of $F(p)$  are at $p = - iq_n$,
where $q_n$ are the real roots of eq.\,\eqref{leading-order-disp}.
 We let $N$ denote
the number of these imaginary zeros;
for $k \neq 0$, $N$ is finite.
We recall that all $q_n$ are positive; hence all $N$ imaginary
zeros of $F(p)$ are 
on the negative imaginary axis, and all $N$ zeros of $F(-p)$ on the positive 
imaginary axis.

Let $\kappa$ and $q$ be the real and imaginary part of $p$: $p = \kappa
+iq$. We let $\mathcal{D}$ denote the rectangular region in the
complex-$p$ plane bounded by the vertical lines $\kappa = \pm \varepsilon$
and horizontal lines 
$q = \varepsilon^{1/2}$ and $q = Q$, where $\varepsilon$ is small and
$Q$ is large enough for the region to contain all $N$ zeros of $F(-p)$ on the
positive imaginary axis. By the argument principle, the total number
of (complex) zeros of the function $F(-p)$ in the region $\mathcal{D}$ is given by
$(2 \pi)^{-1}$ times 
the variation of its argument (i.e., its phase) along the boundary of $\mathcal{D}$. 
In a similar way, one may count the  number of zeros of the
function $F(p)$ in $\mathcal{D}$.

We have
\begin{equation*}
\tan \arg F(\pm p) \! = \! \frac
{\sinh\kappa\sin q \mp \tan k(\sinh\kappa\cos q - \kappa)}
{\cosh\kappa\cos q - 1 \pm \tan k(\cosh\kappa\sin q - q)}.
\end{equation*}
Points on the vertical line $\kappa = \varepsilon$
satisfy
\begin{equation}
\tan \arg F(\pm p) \approx {-\varepsilon} \frac{d}{dq} \ln
|1-\cos q  \pm \tan k(q-\sin q )|.
\label{B3}
\end{equation}
Consider, first, the case of the bottom sign in eq.\,\eqref{B3}. 
The expression between the bars in \eqref{B3}
crosses through zero $N$ times as the line $\kappa = \varepsilon$ is
traced out.  Each time 
zero is crossed, the logarithmic derivative in \eqref{B3} jumps from $-\infty$ to $+\infty$,
 and the argument of $F(-p)$ increases by $\pi$ as we move from one crossing
to the next one.
 The net increase of the argument
as the line $\kappa = \varepsilon$
is traversed from its bottom to the top, is $N \pi$.
In the case of the top sign in \eqref{B3}, on the other hand, the expression in $|\ldots|$ never crosses 
through zero and hence the total increment of the argument
of $F(p)$ is zero. 

 As we move along the vertical
line  $\kappa = -\varepsilon$ from top to bottom, the argument of $F(-p)$ 
increases by another $N \pi$ while the argument of $F(p)$ 
does not acquire any increment. Since there are no zero crossings on the horizontal segments,
the net change of 
the argument of $F(-p)$ along the boundary of $\mathcal{D}$  is
$2N\pi$ while the total increment of ${\rm arg\/} \, F(p)$ is zero.
Therefore $F(-p)$  has only $N$ zeros in the region $\mathcal{D}$ (and they all
lie on the imaginary axis) while the function $F(p)$
has no zeros in $\mathcal{D}$.  This implies that complex singularities 
of $U(p)$ and $W(p)$ cannot accumulate to
the positive imaginary axis.


\begin{thebibliography}{22}

\bibitem{braun}
O. Braun and Yu. S. Kivshar,
Phys. Rep. \textbf{306}, 1 (1998).

\bibitem{hennig}
D. Hennig and G. P. Tsironis
Phys. Rep. \textbf{307}, 333 (1999).

\bibitem{scott}
A. Scott, \emph{Nonlinear Science: emergence and dynamics of
coherent structures} (Oxford University Press, 1999).


\bibitem{christodoulides}
D. N. Christodoulides, F. Lederer, and Y. Silberberg,
Nature \textbf{424}, 817 (2003).

\bibitem{campbell}
D. K. Campbell, S. Flach, and Yu. S. Kivshar,
Phys. Today \textbf{57}(1), 43 (2004).


\bibitem{trambettoni}
A. Trombettoni and A. Smerzi,
Phys. Rev. Lett. \textbf{86}, 2353 (2001).


\bibitem{Feddersen}
H. Feddersen, in \emph{Nonlinear Coherent Structures in Physics
and Biology. Proceedings of the 7th Interdisciplinary
Workshop Held at Dijon, France, 4-6 June 1991.}
Eds. M. Remoissenet and M. Peyrard, Lecture Notes
in Physics vol. 393 (Springer, Berlin, 1991), pp. 159--167.

\bibitem{duncan}
D. B. Duncan, J. C. Eilbeck, H. Feddersen, and J. A. D. Wattis,
Physica D \textbf{68}, 1 (1993).

\bibitem{flach}
S. Flach and C. R. Willis,
Phys. Rep. \textbf{295}, 181 (1998).

\bibitem{Flach_Kladko}
S. Flach and K. Kladko,
Physica D \textbf{127}, 61 (1999).

\bibitem{FKZ}
S. Flach, Y. Zolotaryuk and K. Kladko,
Phys. Rev. E \textbf{59}, 6105 (1999).

\bibitem{kevrekidis}
P. G. Kevrekidis, K. \O{}. Rasmussen, and A. R. Bishop,
Int. J. Mod. Phys. B \textbf{15}, 2833 (2001).

\bibitem{ablowitz}
M. J. Ablowitz, Z. H. Musslimani, and G. Biondini,
Phys. Rev. E \textbf{65}, 026602 (2002).

\bibitem{Kundu} K. Kundu, J. Phys. A: Math. Gen. {\bf 35}, 8109 (2002).

\bibitem{eilbeck}
J. C. Eilbeck and M. Johansson,
in \emph{Proceedings of the Third Conference on Localization
  and Energy Transfer in Nonlinear Systems, San Lorenzo de El Escorial, Madrid,
  Spain, June 17--21, 2002} (World Scientific, Singapore, 2003), pp. 44--67.

\bibitem{Malomed_collisions_1}
I. E. Papacharalampous,  P. G. Kevrekidis, B. A. Malomed, and D. J. Frantzeskakis,
  Phys. Rev. E \textbf{68}, 046604 (2003).

\bibitem{Malomed_collisions_2}
 S. V. Dmitriev, P. G. Kevrekidis, B. A. Malomed, and D. J. Frantzeskakis,
  Phys. Rev. E \textbf{68}, 056603 (2003).

\bibitem{pelinovsky}
D. E. Pelinovsky and V. M. Rothos,
Physica D \textbf{202}, 16 (2005).

\bibitem{Malomed_nonlinearity_management} 
J. Cuevas, B. A. Malomed, and P. G. Kevrekidis, Phys. Rev. E {\bf 71}, 066614 (2005).


\bibitem{gomez1}
J. G\'omez-Garde\~nez, F. Falo, and L. M. Flor\'ia,
Phys. Lett. A \textbf{332}, 213 (2004).

\bibitem{gomez}
J. G\'omez-Garde\~nez, L. M. Flor\'ia, M. Peyrard, and A. R. Bishop,
Chaos \textbf{14}, 1130 (2004).


\bibitem{efremidis}
N. K. Efremidis, S. Sears, D. N. Christodoulides, J. W. Fleischer, and M. Segev,
Phys. Rev. E \textbf{66}, 046602 (2002).

\bibitem{fleischer}
J. W. Fleischer, T. Carmon, M. Segev, N. K. Efremidis, and D. N. Christodoulides,
Phys. Rev. Lett. \textbf{90}, 023902 (2003).

\bibitem{chen}
F. Chen, C. E. R\"uter, D. Runde, D. Kip, V. Shandarov, O. Manela, and M. Segev,
Opt. Exp. \textbf{13}, 4314 (2005).


\bibitem{fleischer-nature}
J. W. Fleischer, M. Segev, N. K. Efremidis, and D. N. Christodoulides,
Nature \textbf{422}, 147 (2003).

\bibitem{hadzievski}
L. Had\v{z}ievski, A. Maluckov, M. Stepi\'c, and D. Kip,
Phys. Rev. Lett. \textbf{93}, 033901 (2004).

\bibitem{stepic}
M. Stepi\'c, D. Kip, L. Had\v{z}ievski, and A. Maluckov,
Phys. Rev. E \textbf{69}, 066618 (2004).

\bibitem{cuevas}
J. Cuevas and J. C. Eilbeck,
Phys. Lett. A \textbf{358}, 15 (2006).

\bibitem{vicencio}
R. A. Vicencio and M. Johansson,
Phys. Rev. E \textbf{73}, 046602 (2006).


\bibitem{gatz1}
S. Gatz and J. Herrmann,
J. Opt. Soc. Am. B \textbf{8}, 2296 (1991).

\bibitem{gatz2}
S. Gatz and J. Herrmann,
Opt. Lett. \textbf{17}, 484 (1992).


\bibitem{tikhonenko}
V. Tikhonenko, J. Christou, and B. Luther-Davies,
Phys. Rev. Lett. \textbf{76}, 2698 (1996).


\bibitem{vidal}
F. Vidal and T. W. Johnston,
Phys. Rev. E \textbf{55}, 3571 (1997).


\bibitem{khare}
A. Khare, K. \O{}. Rasmussen, M. R. Samuelsen, and A. Saxena,
J. Phys. A: Math. Gen. \textbf{38}, 807 (2005).


\bibitem{BOP} I. V. Barashenkov, O. F. Oxtoby, and
D. E. Pelinovsky, Phys. Rev. E {\bf 72} 035602(R) (2005).


\bibitem{OPB}
O. F. Oxtoby, D. E. Pelinovsky, and I. V. Barashenkov,
Nonlinearity \textbf{19}, 217 (2006).


\bibitem{vinetskii}
V. O. Vinetskii and N. V. Kukhtarev,
Sov. Phys. Solid State \textbf{16}, 2414 (1975).


\bibitem{Lindstedt} See e.g. A. H. Nayfeh and D. T. Mook, Nonlinear
Oscillations (J. Wiley, New York, 1979); D. W. Jordan and P. Smith, Nonlinear
Ordinary Differential Equations (Oxford University Press, 1999).


\bibitem{AL} M.J. Ablowitz and J.F. Ladik, Stud. Appl. Math.
{\bf 55} 213 (1976); J. Math. Phys. {\bf 17} 10011 (1976).


\bibitem{Laedke_Kluth_Spatschek}
E. W. Laedke, O. Kluth, and K. H. Spatschek, Phys. Rev.
{\bf E 54} 4299 (1996).


\bibitem{pelinovsky-exceptional}
D. E. Pelinovsky, Nonlinearity \textbf{19}, 2695 (2006).


\bibitem{eleonskii}
V. M. Eleonksii, N. E. Kulagin, N. S. Novozhilova, and V. P. Silin,
Teor. Mat. Fiz. \textbf{60}, 395 (1984).


\bibitem{sk}
H. Segur and M. D. Kruskal,
Phys. Rev. Lett. \textbf{58}, 747 (1987).

\bibitem{sk2}
M. D. Kruskal and H. Segur,
Stud. Appl. Math. \textbf{85}, 129 (1991).


\bibitem{pomeau}
Y. Pomeau, A. Ramani, and B. Grammaticos,
Physica D \textbf{31}, 127 (1988).


\bibitem{GrimshawJoshi}
R. H. J. Grimshaw and N. Joshi,
SIAM J. Appl. Math. \textbf{55}, 124 (1995).

\bibitem{GrimshawNLS}
R. H. J. Grimshaw,
Stud. Appl. Math. \textbf{94}, 257 (1995).


\bibitem{Tovbis1}
A. Tovbis, M. Tsuchiya, and C. Jaff\'e,
Chaos \textbf{8}, 665 (1998).

\bibitem{Tovbis2}
A. Tovbis,
Contemporary Mathematics \textbf{255}, 199 (2000).

\bibitem{Tovbis3}
A. Tovbis, Stud. Appl. Math. \textbf{104}, 353 (2000).

\bibitem{TovPel}
A. Tovbis and D. Pelinovsky,
Nonlinearity \textbf{19}, 2277 (2006).


\bibitem{embedded_review}
J. Fujioka, A. Espinosa-Cer\'on, and R. F. Rodr\'iguez,
Rev. Mex. F\'is. \textbf{52}, 6 (2006).


\bibitem{Malomed_embedded_0}
S. Gonz\'alez-P\'erez-Sandi, J. Fujioka, and B. A. Malomed,
Physica D \textbf{197}, 86 (2004).


\bibitem{Malomed_embedded}
B. A. Malomed, J. Fujioka, A. Espinosa-Cer\'on, R. F. Rodr\'iguez, and
S. Gonz\'alez,
Chaos \textbf{16}, 013112 (2006).


\bibitem{Malomed_embedded_2}
K. Yagasaki, A. R. Champneys, and B. A. Malomed,
Nonlinearity {\bf 18}, 2591 (2005).


\bibitem{melvin}
T. R. O. Melvin, A. R. Champneys, P. G. Kevrekidis, and J. Cuevas,
Phys. Rev. Lett. \textbf{97}, 124101 (2006).


\end{thebibliography}
\end{document}